\documentstyle[sprocl]{article}
\def\lsim{\mathrel{\rlap{\lower4pt\hbox{\hskip1pt$\sim$}}
    \raise1pt\hbox{$<$}}}         
\def\gsim{\mathrel{\rlap{\lower4pt\hbox{\hskip1pt$\sim$}}
    \raise1pt\hbox{$>$}}}         

\input psfig.sty

\def\Journal#1#2#3#4{{#1} {\bf #2}, #3 (#4)}

\def\NPB{{\em Nucl. Phys.} B}
\def\PLB{{\em Phys. Lett.}  B}
\def\PRL{\em Phys. Rev. Lett.}
\def\PRD{{\em Phys. Rev.} D}
\def\PRC{{\em Phys. Rev.} C}

\def\ARAA{\em Ann. Rev. Astron. Astrophys.}
\def\AJ{\em Ap. J.}
\def\AJS{\em Ap. J. Suppl.}
\def\RMP{\em Rev. Mod. Phys.}
\def\N{\em Nature}
\def\SJNP{\em Sov. J. Nucl. Phys.}
\def\PPNP{\em Prog. Part. Nucl. Phys.}
\def\AA{\em Astron. Astrophys.}

\def\be{\begin{equation}}
\def\ee{\end{equation}}
\def\bea{\begin{eqnarray}}
\def\eea{\end{eqnarray}}

\begin{document}

\title{TOPICS IN NEUTRINO ASTROPHYSICS}

\author{W. C. HAXTON}

\address{Institute for Nuclear Theory, Box 351550, and
Department of Physics, Box 351560\\
University of Washington, Seattle, WA 98195, USA\\
E-mail: haxton@phys.washington.edu}


\maketitle\abstracts{ In these TASI summer school lectures I discuss three topics in
neutrino astrophysics: the solar neutrino problem, stellar 
cooling by neutrino emission, and the role of neutrinos in the
nucleosynthesis that occurs within core-collapse supernovae.}

\section{Introduction}
Part of the interest in neutrino astrophysics has to do with
the fascinating interplay between nuclear and particle physics issues ---
e.g., whether neutrinos are massive and undergo flavor oscillations,
whether they have detectable electromagnetic moments, etc. ---
and astrophysical phenomena, such as the clustering of matter
on large scales, the mechanisms responsible for the synthesis
of nuclei, and the evolution of stars.  The three lectures here
are intended to illustrate this interplay.  The first lecture
reviews the solar neutrino problem which, along with
the atmospheric neutrino problem, provides perhaps our
strongest direct evidence that new physics lurks beyond the
standard model.  The second has to with the implications of 
neutrino properties --- e.g., whether neutrinos are Dirac or Majorana
particles --- for stellar cooling.  The final lecture describes
the nucleosynthesis we think accompanies a supernova explosion,
and why that synthesis is a delicate probe of neutrino
oscillations. 

These lectures as well of those from several other 1998 TASI 
speakers share a common subtheme: how the extraordinary technical
revolution in astronomy and astrophysics has made  
the microphysics of the universe more relevant.  It is the
precise data coming from the new generation of great observatories ---
maps of the cosmic microwave spectrum, precision measurements 
of the products of big bang nucleosynthesis, measurements of
the solar neutrino spectrum, Hubble Space Telescope (HST) 
abundance distributions from early, metal-poor stars,
detection of gamma ray bursts from cosmological sources ---
that allow us to form the connections between observations and
the underlying microphysics.  This is the driving force that
is making the field of nuclear and particle astrophysics of
such interest to both senior physicists and new students entering
the field. 
  
\section{Solar Neutrinos~\protect\cite{haxtonsn}}
More than three decades ago Ray Davis, Jr. and his collaborators~\cite{davis}
constructed a 0.615 kiloton C$_2$Cl$_4$ radiochemical solar
neutrino detector in the Homestake Gold Mine, one mile beneath
Lead, South Dakota.  Within a few years it was apparent that 
the number of neutrinos detected was considerably below the
predictions of the standard solar model, that is, the standard
theory of main sequence stellar evolution. 

Today the results from the $^{37}$Cl detector, which have become
quite accurate due to 30 years of careful measurement, have
been augmented by results from four other experiments, the
SAGE~\cite{sage} and GALLEX~\cite{gallex} gallium experiments and the Kamiokanda~\cite{k} and
SuperKamiokande~\cite{sk} water Cerenkov detectors.  It now appears that
the combined results are very difficult to explain --- some have argued
impossible --- by any plausible change in the standard solar
model (SSM).  Thus most believe that the answer to the solar neutrino
problem is new particle physics, most likely some effect
like solar neutrino oscillations associated with massive
neutrinos.  With the recent news that SuperKamiokande sees
direct evidence for $\nu_\mu$ oscillations in the azimuthal
dependence of atmospheric~\cite{atmos} neutrinos, it seems that we may be
on the threshold of a major discovery.

The purpose of this first (and longest) lecture is to summarize
the solar neutrino problem and the arguments that it represents
new particle physics.

\subsection{The Standard Solar Model~\protect\cite{bbp98}}
Solar models trace the evolution of the sun over the past
4.6 billion years of main sequence burning, thereby predicting
the present-day temperature and composition profiles of the solar
core that govern neutrino production.  Standard solar models 
share four basic assumptions:
    
\noindent
* The sun evolves in hydrostatic equilibrium, maintaining
a local balance between the gravitational force and the pressure
gradient.  To describe this condition in detail, one must 
specify the equation of state as a function of temperature,
density, and composition.

\noindent
* Energy is transported by radiation and convection.  While
the solar envelope is convective, radiative transport dominates
in the core region where thermonuclear reactions take place.
The opacity depends sensitively on the solar composition, particularly
the abundances of heavier elements.

\noindent
* Thermonuclear reaction chains generate solar energy.
The standard model predicts that over 98\% of this energy
is produced from the pp chain conversion of four protons into $^4$He
(see Fig. 1)
\begin{equation}
          4p \rightarrow ^4\mathrm{He} + 2e^+  + 2 \nu_e 
\end{equation}
with proton burning through the CNO cycle contributing the remaining 2\%.  The sun is
a large but slow reactor: the core temperature, $T_c \sim  1.5 \cdot 10^7$ K,
results in typical center-of-mass energies for reacting particles
of $\sim$ 10 keV, much less than the Coulomb barriers inhibiting
charged particle nuclear reactions.  Thus reaction cross
sections are small: in most cases, as laboratory measurements are
only possible at higher energies, cross section data must be
extrapolated to the solar energies of
interest.
    
\noindent
* The model is constrained to produce today's solar
radius, mass, and luminosity.  An important assumption of
the standard model is that the sun was highly convective,
and therefore uniform in composition, when it first
entered the main sequence.  It is furthermore assumed
that the surface abundances of metals (nuclei with A $>$ 5)
were undisturbed by the subsequent evolution, and thus
provide a record of the initial solar metallicity.  The
remaining parameter is the initial $^4$He/H ratio, which
is adjusted until the model reproduces the present solar
luminosity after 4.6 billion years of evolution.  The resulting
$^4$He/H mass fraction ratio is typically 0.27 $\pm$ 0.01,
which can be compared to the big-bang value of 0.23 $\pm$ 0.01.  
Note that the sun was formed from previously processed
material.

\begin{figure}[htb]
\psfig{bbllx=0.5cm,bblly=4.0cm,bburx=18cm,bbury=18.5cm,figure=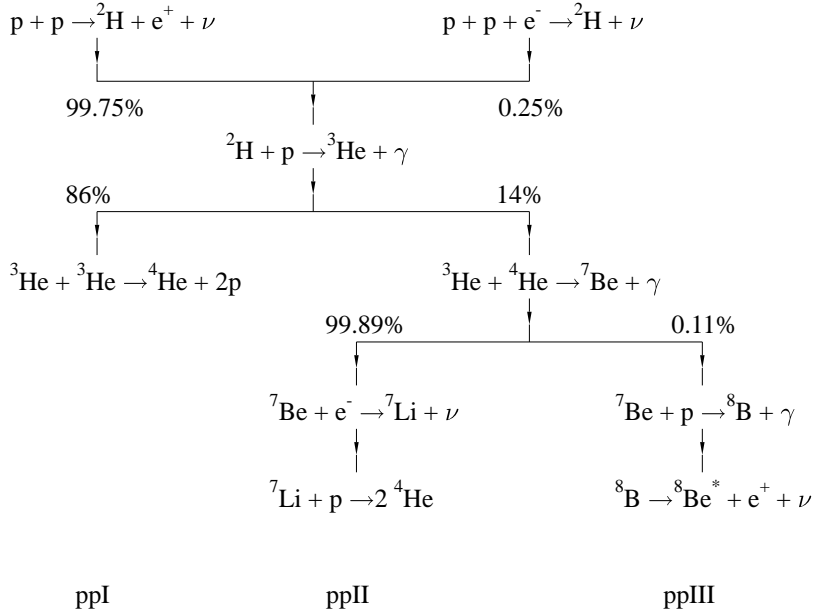,height=3.3in}
\caption{The solar pp chain.}
\end{figure}
  
The model that emerges is an evolving sun.  As the
core's chemical composition changes, the opacity and
core temperature rise, producing a 44\% luminosity increase
since the onset of the main sequence.
The temperature rise governs the competition between the
three cycles of the pp chain: the ppI cycle dominates below
about 1.6 $\cdot 10^7$ K; the ppII cycle between
(1.7-2.3) $\cdot 10^7$K; and the ppIII above 2.4 $\cdot 10^7$K.
The central core temperature of today's SSM is about 
1.55 $\cdot 10^7$K.

The competition between the cycles determines the pattern
of neutrino fluxes.  Thus one consequence of the thermal
evolution of our sun is that the $^8$B neutrino
flux, the most temperature-dependent component, proves to
be of relatively recent origin: the predicted flux 
increases exponentially with a doubling period of about
0.9 billion years.

A final aspect of SSM evolution is the formation of composition
gradients on nuclear burning timescales.  Clearly there is a
gradual enrichment of the solar core in $^4$He, the ashes 
of the pp chain.  Another element, $^3$He, is a sort of
catalyst for the pp chain, being produced and then consumed,
and thus eventually reaching some equilibrium abundance.
The timescale for equilibrium to be established as well
as the eventually equilibrium abundance are both sharply
decreasing functions of temperature, and thus increasing
functions of the distance from the center of the core.
Thus a steep $^3$He density gradient is
established over time.

The SSM has had some notable successes.  From helioseismology
the sound speed profile $c(r)$ has been very accurately 
determined for the outer 90\% of the sun, and is in excellent
agreement with the SSM.  Such studies verify important predictions
of the SSM, such as the depth of the convective zone.
However the SSM is not a complete model in that it does 
not explain all features of solar structure, such as the 
depletion of surface Li by two orders of magnitude.  This is 
usually attributed to convective processes that operated at
some epoch in our sun's history,
dredging Li to a depth where burning takes place.
  
The principal neutrino-producing reactions of the pp chain
and CNO cycle are summarized in Table 1.  The first six reactions
produce $\beta$ decay neutrino spectra having allowed shapes with 
endpoints given by E$_\nu^{\rm max}$.  Deviations from an allowed spectrum
occur for $^8$B neutrinos because the $^8$Be final state is a broad 
resonance.  The last two reactions
produce line sources of electron capture neutrinos, with
widths $\sim$ 2 keV characteristic of the temperature of the solar core.
Measurements of the  pp, $^7$Be, and $^8$B neutrino fluxes will
determine the relative contributions of the  ppI,  ppII, and
ppIII cycles to solar energy generation.  As discussed above, 
and as later illustrations will show more clearly, 
this competition is governed in large
classes of solar models by a single parameter, the central
temperature $T_c$.  The flux predictions of 
the 1998 calculations of Bahcall, Basu, and Pinsonneault~\cite{bbp98} (BP98) 
and of Brun, Turck-Chieze and Morel~\cite{tcl} are included
in Table 1.

\begin{table}[t]
\caption{Solar neutrino sources and the flux predictions of the
BP98 and Brun/Turck-Chieze/Morel SSMs in cm$^{-2}$s$^{-1}$.}
\vspace{0.2cm}
\begin{center}
\begin{tabular}{|c|c|c|c|}
\hline
 & & & \\
Source & E$_\nu^{max}$ (MeV) & BP98 & BTCM98 \\
& & & \\
\hline
& & & \\
p + p $\rightarrow ^2$H + e$^+ + \nu$ & 0.42 & 5.94E10 & 5.98E10 \\
$^{13}$N $\rightarrow ^{13}$C + e$^+ + \nu$ & 1.20 & 6.05E8 & 4.66E8 \\
$^{15}$O $\rightarrow ^{15}$N + e$^+ + \nu$ & 1.73 & 5.32E8 & 3.97E8 \\
$^{17}$F $\rightarrow ^{17}$O + e$^+ + \nu$ & 1.74 & 6.33E6 & \\
$^8$B $\rightarrow ^8$Be + e$^+ + \nu$ & $\sim$ 15 & 5.15E6 & 4.82E6 \\
$^3$He + p $\rightarrow ^4$He + e$^+ + \nu$ & 18.77 & 2.10E3 & \\
$^7$Be + e$^- \rightarrow ^7$Li + $\nu$ & 0.86 (90\%) & 4.80E9 & 4.70E9 \\
 & 0.38 (10\%) & & \\
p + e$^-$ + p $\rightarrow ^2$H + $\nu$ & 1.44 & 1.39E8 & 1.41E8 \\
 & & & \\
\hline
\end{tabular}
\end{center}
\end{table}
  
\subsection{Solar Neutrino Detection~\protect\cite{haxtontm}}
Let us start with a brief reminder about low energy neutrino-nucleus interactions
in detectors.  Consider the charged current reaction 
\begin{equation}
\nu_e + (A,Z) \rightarrow e^- + (A,Z+1)
\end{equation}
Because the momentum transfer to the nucleus is very small for
solar neutrinos, it can be neglected in the weak propagator,
leading to an effective contact current-current interaction.
If we begin with the simplest (though fictitious) case of
the free neutron decay $n \rightarrow p$, the corresponding transition amplitude
is then
\begin{equation}
S_{fi} = {G_F \over \sqrt{2}} \cos \theta_C 
\bar{u}(p) \gamma_\mu (1 - g_A \gamma_5) u(n)
\bar{u}(e) \gamma^\mu (1 - \gamma_5) u(\nu)
\end{equation}
where $G_F$ is the weak coupling constant measured in muon decay
and $\cos \theta_c$ gives the amplitude for the weak interaction
to connect the u quark to its first-generation partner, the d quark.
The origin of this effective amplitude is the underlying standard
model predictions for the elementary quark and lepton currents.
The weak interactions at this level are predicted by the standard
model to be exactly left handed.  Experiment shows that the
effective coupling of the W boson to the nucleon is governed by
$\gamma_\mu (1 - g_A \gamma_5)$, as noted above,
where $g_A \sim 1.26$.
The axial coupling is thus shifted from its underlying value
by the strong interactions responsible for the binding of the
quarks within the nucleon.

If an isolated nucleon were the target, one could proceed to
calculate the cross section from the effective nucleon 
current given above.
The extension to nuclear systems traditionally begins with the
observation that nucleons in the nucleus are rather nonrelativistic,
$v/c \sim 0.1$.  The amplitude 
$\bar{u}(p) \gamma^\mu(1-g_A \gamma_5) u(n)$ can be expanded
in powers of $p/M$.
The leading vector and axial operators
are readily found to be
\begin{eqnarray}
 \gamma_0&:&~~~1 \nonumber \\
\vec{\gamma}&:&~~~\vec{p}/M \sim v/c \nonumber \\
\gamma_0 \gamma_5&:&~~~\vec{\sigma} \cdot \vec{p}/M \sim v/c \nonumber \\
\vec{\gamma} \gamma_5&:&~~~\vec{\sigma} \nonumber
\end{eqnarray}
Thus it is the time-like part of the vector current and the 
space-like part of the axial-vector current that survive in the
nonrelativistic limit.

(In a nucleus these currents must be corrected for the presence
of meson exchange contributions.  The corrections to the 
vector charge and axial three-current, which we just pointed out
survive in the nonrelativistic limit, are of order $(v/c)^2
\sim$ 1\%.  Thus the naive one-body currents
are a very good approximation to the nuclear currents.  In
contrast, exchange current corrections to the axial charge and
vector three-current operators are of order $v/c$, and thus
of relative order 1.  This difficulty for the vector three-current
can be largely circumvented, because 
current conservation as embodied in the generalized Siegert's
theorem allows one to rewrite important parts of this
operator in terms of the
vector charge operator.  In the long-wavelength limit 
appropriate to solar neutrinos, all terms unconstrained by
current conservation do not survive.  In effect, one has
replaced a current operator with large two-body corrections
by a charge operator with only small corrections.
In contrast, the axial charge operator is significantly altered
by exchange currents even for long-wavelength processes like
$\beta$ decay.  Typical axial-charge $\beta$ decay rates are
enhanced by $\sim$ 2 because of exchange currents.)
  
If such a nonrelativistic reduction is done for our single
current one obtains
\begin{eqnarray}
S_{fi}  & \sim &  \cos \theta_c {G_F \over \sqrt{2}} 
( \phi^\dagger (p) \phi(n) \bar{u}(e) \gamma^0(1 -\gamma_5)u(\nu) \nonumber \\
  & & - \phi^\dagger (p) g_A \vec{\sigma} \phi(n) \cdot \bar{u}(e)
\vec{\gamma}(1-\gamma_5)u(\nu) ) 
\end{eqnarray}
where the $\phi$ are now two-component Pauli spinors for the nucleons.
The above result can be generalized to include $\bar{\nu}_e$ 
reactions 
by introducing the isospin operators
$\tau_\pm$ where $\tau_+$ $\mid$ n$\rangle$ = $\mid$ p$\rangle$ and $\tau_-$$\mid$ p$\rangle$ = $\mid$ n$\rangle$, with
all other matrix elements being zero.  Thus
\[ \phi^\dagger (p) \phi(n) \rightarrow \phi^\dagger (N) \tau_\pm 
\phi(N) \]
\[ \phi^\dagger (p) \vec{\sigma} \phi(n) \rightarrow \phi^\dagger (N)
\vec{\sigma} \tau_\pm \phi(N). \]
This result easily generalizes to nuclear decay.  Given our
comments about exchange currents, the first step is the
replacement
\[ \tau_\pm \rightarrow \sum_{i=1}^A \tau_\pm(i) \]
\[ \sigma \tau_\pm \rightarrow \sum_{i=1}^A \sigma(i)
\tau_\pm(i). \]
Plugging $S_{fi}$ into the standard cross section formula
(which involves an average over initial and sum over final
nuclear spins of the square of the transition amplitude)
then yields the allowed nuclear matrix element
\begin{equation}
{1 \over 2J_i+1} (|\langle f || \sum_{i=1}^A \tau_\pm (i) || i \rangle |^2
+ g_A^2 |\langle f || \sum_{i=1}^A \sigma(i) \tau_\pm(i) || i \rangle|^2).
\end{equation}

Our initial calculation for the nucleon treated that particle
as structureless.  Implicitly we assumed that the momentum transfer is 
much smaller than the inverse nucleon size.  If we take 10 MeV
as a typical solar neutrino momentum transfer, these quantities
would be in the ratio 1:20.  For a light nucleus, the 
corresponding result might be 1:10.  This long-wavelength
approximation in combination with the nonrelativistic 
approximation yields the allowed result, where only Fermi
and Gamow-Teller operators survive.  These are the 
spin-independent and spin-dependent operators appearing above.
  
The Fermi operator is proportional to the isospin raising/lowering operator:
in the limit of good isopsin, which typically is good to 5\% or
better in the description of low-lying nuclear states,
it can only connect states in the same isospin multiplet,
that is, states with a common spin-spatial structure.
If the initial state has isospin $(T_i, M_{Ti})$, this final
state has $(T_i, M_{Ti} \pm 1)$ for $(\nu,e^-)$ and
$(\bar{\nu},e^+)$ reactions, respectively, and is called the isospin analog state (IAS).
In the limit
of good isospin the sum rule for this operator in then
particularly simple
\begin{equation}
\sum_f {1 \over 2J_i+1} | \langle f || \sum_{i=1}^A \tau_+(i) || i \rangle |^2 =
{1 \over 2J_i+1} | \langle IAS || \sum_{i=1}^A \tau_+(i) || i \rangle |^2 = |N-Z|. 
\end{equation}
The excitation energy of the IAS relative to the parent ground
state can be estimated accurately from the Coulomb energy
difference 
\begin{equation}
E_{IAS} \sim ({1.728 Z \over 1.12A^{1/3} + 0.78} - 1.293) \mathrm{MeV}. 
\end{equation}
The angular distribution of the outgoing electron for a pure
Fermi $(N,Z) + \nu \rightarrow (N-1,Z+1) + e^-$ transition is 1 + $\beta \cos \theta_{\nu e}$, 
and thus forward peaked.  Here $\beta$ is the electron velocity.

The Gamow-Teller (GT) response is more complicated, as the 
operator can connect the ground state to many states in the
final nucleus.  In general we do not have a precise probe of 
the nuclear GT response apart from weak interactions themselves.
However a good approximate probe is provided by forward-angle
(p,n) scattering off nuclei, a technique that has been 
developed in particular by experimentalists at the Indiana
University Cyclotron Facility.  The (p,n) reaction transfers
isospin and thus is superficially like $(\nu,e^-)$.  At
forward angles (p,n) reactions
involve negligible three-momentum transfers to the nucleus.
Thus the nucleus should not be radially excited.  It thus
seems quite plausible that forward-angle (p,n) reactions
probe the isospin and spin of the nucleus, the macroscopic 
quantum numbers, and thus the Fermi and GT responses.
For typical transitions, the correspondence between (p,n) and
the weak GT operators is believed to be accurate to about 10\%.
Of course, in a specific transition, much larger discrepancies
can arise.

The (p,n) studies demonstrate that the GT strength tends to 
concentrate in a broad resonance 
centered at a position $\delta = E_{GT} - E_{IAS}$ relative
to the IAS given by 
\begin{equation}
 \delta \sim (7.0 -28.9 {N-Z \over A})~\mathrm{MeV}. 
\end{equation}
Thus while the peak of the GT resonance is substantially above the IAS for
$N \sim Z$ nuclei, it drops with increasing neutron excess.
Thus $\delta \sim 0$ for Pb.  A typical value for the full
width at half maximum $\Gamma$ is $\sim$ 5 MeV.

The approximate Ikeda sum rule constrains the difference
in the $(\nu,e^-)$ and $(\bar{\nu},e^+)$ strengths
\begin{equation}
\sum_f ( |M_{GT}^{fi}(\nu,e^-)|^2 - |M_{GT}^{fi}(\bar{\nu},e^+)|^2 )
= 3(N-Z)
\end{equation}
where
\begin{equation}
|M_{GT}^{fi}(\nu,e^-)|^2 = {1 \over 2J_i+1} 
|\langle f || \sum_{i=1}^A \sigma (i) \tau_+(i) || i \rangle |^2. 
\end{equation}
In many cases of interest in heavy nuclei, the strength in the
$(\bar{\nu},e^+)$ direction is largely blocked.  For example, in a
naive $2s1d$ shell model description of $^{37}$Cl, 
the p $\rightarrow$ n direction is blocked by the closed
neutron shell at N=20.  Thus this relation can provide an 
estimate of the total $\beta^-$ strength.  Experiment shows
that the $\beta^-$ strength found in and below the GT 
resonance does not saturate the Ikeda sum rule, typically 
accounting for $\sim (60-70)$ \% of the total.  Measured and
shell model predictions of individual GT transition strengths
tend to differ systematically by about the same factor.
Presumably the missing strength is spread over a broad interval
of energies above the GT resonance.  This is not unexpected
if one keeps in mind that the shell model is an approximate
effective theory designed to describe the long wavelength modes
of nuclei: such a model should require effective operators,
renormalized from their bare values.  Phenomenologically, the
shell model seems to require~\cite{brown} $g_A^{eff} \sim$ 1.0 as well as
a small spin-tensor term $(\sigma \otimes Y_2(\hat{r}) )_{J=1}$
of relative strength $\sim$ 0.1.

The angular distribution of GT $(N,Z) + \nu_e \rightarrow 
(N-1,Z+1) + e^-$ reactions is $3 - \beta \cos \theta_{\nu e}$,
corresponding to a gentle peaking in the backward direction.
 
The above discussion of allowed responses can be repeated for
neutral current processes such as $(\nu,\nu')$.  The analog
of the Fermi operator contributes only to elastic processes,
where the standard model nuclear weak charge is approximately
the neutron number.  As this operator does not generate
transitions, it is not yet of much interest for 
solar or supernova neutrino detection, though there are efforts
to develop low-threshold detectors (e.g., cryogenic technologies)
for recording the modest nuclear recoil energies.
The analog of the GT response involves
\begin{equation}
|M_{GT}^{fi}(\nu,\nu')|^2 = {1 \over 2J_i+1}
|\langle f || \sum_{i=1}^A \sigma(i) {\tau_3(i) \over 2} || i
\rangle |^2. 
\end{equation}
The operator appearing in this expression is familiar from
magnetic moments and magnetic transitions, where the 
large isovector magnetic moment ($\mu_v \sim$ 4.706) often
leads to it dominating the orbital and isoscalar spin operators.

Finally, there is one purely leptonic reaction of great interest,
since it is the reaction exploited by Kamiokande and 
SuperKamiokande.  Electron neutrinos can scattered off electrons
via both charged and neutral current reactions.  The cross
section calculation is straight forward and will not be 
repeated here.  Two features of the result are of importance
for our later discussions, however.  Because of the neutral 
current contribution, 
heavy-flavor $(\nu_\mu$ and $\nu_\tau)$ also scatter off electrons, but with a 
cross section reduced by about a factor of seven at low
energies.  Second, for neutrino energies well above the electron 
rest mass, the scattering is sharply forward peaked.  Thus
this reaction allows one to exploit the position of the sun
in separating the solar neutrino signal from 
a large but isotropic background.
  
As I mentioned earlier, the first experiment performed was
one exploiting the reaction
\[ ^{37}\mathrm{Cl}(\nu,e^-)^{37}\mathrm{Ar}. \]
As the threshold for this reaction is 0.814 MeV, the important
neutrino sources are the $^7$Be and $^8$B reactions.
The $^7$Be neutrinos excite just the GT transition to
the ground state, the strength of which is known from the 
electron capture lifetime of $^{37}$Ar.  The $^8$B neutrinos
can excite all bound states in $^{37}$Ar, including the
dominant transition to the IAS residing at an excitation
of 4.99 MeV.  The strength of excite-state GT transitions
can be determined from the $\beta$ decay $^{37}$Ca$(\beta^+)^{37}$K,
which is the isospin mirror reaction to $^{37}$Cl$(\nu,e^-)^{37}$Ar.
The net result is that, for SSM fluxes, 78\% of the capture
rate should be due to $^8$B neutrinos, and 15\% to $^7$Be
neutrinos.  The measured capture rate~\cite{lande} 2.56 $\pm 0.16 \pm 0.16$ SNU
(1 SNU = 10$^{-36}$ capture/atom/sec) is about 1/3 the 
standard model value.

Similar radiochemical experiments were done
by the SAGE and GALLEX collaborations using a different 
target, $^{71}$Ga.  The special properties of this target include
its low threshold and an unusually strong transition to the ground state of 
$^{71}$Ge, leading to a large pp neutrino cross section (see Fig. 2).
The experimental capture rates are
$66 \pm 13 \pm 6$ and $76 \pm 8$ SNU for the SAGE and GALLEX
detectors, respectively.  The SSM prediction is about 130 
SNU~\cite{bahcallb}.  Most important, since the pp flux is directly constrained
by the solar luminosity in all steady-state models, there is
a minimum theoretical value for the capture rate of 79 SNU,
given standard model weak interaction physics.  Note there
are substantial uncertainties in the $^{71}$Ga cross section
due to $^7$Be neutrino capture to two excited states of unknown 
strength.  These uncertainties were greatly reduced by
direct calibrations of both detectors using $^{51}$Cr
neutrino sources.

\begin{figure}[htb]
\psfig{bbllx=0.0cm,bblly=4.0cm,bburx=16cm,bbury=22.5cm,figure=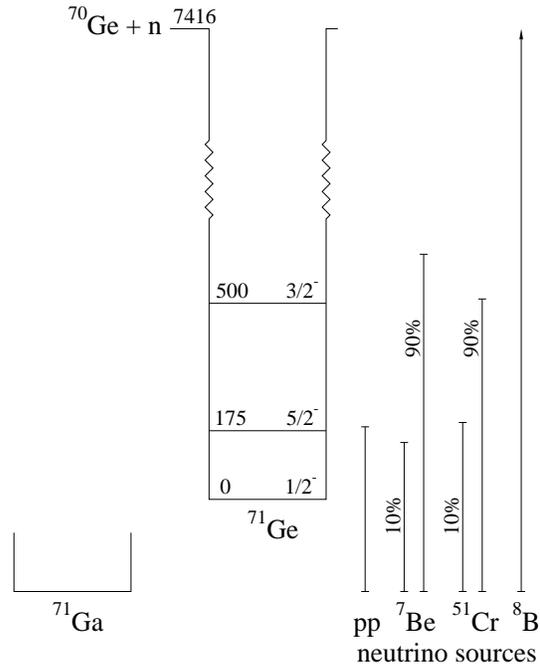,height=3.8in}
\caption{Level scheme for $^{71}$Ge showing the excited states
that contribute to absorption of pp, $^7$Be, $^{51}$Crm and
$^8$B neutrinos.}
\end{figure}
  
The remaining experiments, Kamiokande II/III and SuperKamiokande,
exploited water Cerenkov detectors to view solar
neutrinos event-by-event.  Solar neutrinos scatter off electrons,
with the recoiling electrons producing the Cerenkov radiation
that is then recorded in surrounding phototubes.  
Thresholds are determined by background rates; SuperKamiokande
is currently operating with a trigger at approximately six MeV.
The initial experiment, Kamiokande II/III, found a flux of
$^8$B neutrinos of (2.91 $\pm 0.24 \pm 0.35) \cdot 10^6$/cm$^2$s after
about a decade of measurement.  Its much larger successor
SuperKamiokande, with a 22.5 kiloton fiducial volume,
yielded the result $(2.37 \pm 0.06 \pm 0.08) \cdot 10^6$/cm$^2$s
after the first 374 days of measurements.  This is about
36\% of the SSM flux.  This result continues to improve in accuracy. \\
  
\subsection{Uncertainties in Standard Solar Model Parameters}

\noindent
The pattern of solar neutrino fluxes that has emerged from these
experiments is
\begin{eqnarray}
\phi (pp) & \sim & 0.9 \, \phi^{\rm {SSM}} (pp)\nonumber \\
\phi (^7{\rm {Be}}) & \sim & 0 \nonumber\\
\phi (^8 {\rm B}) & \sim & 0.43 \, \phi^{\rm {SSM}} (^8{\rm B}).  
\end{eqnarray}
A reduced $^8$B neutrino flux can be produced by lowering
the central temperature of the sun somewhat, as $\phi(^8$B)$\sim T_c^{18}$.  However, such
an adjustment, either by varying the parameters of the SSM or by
adopting some nonstandard physics, tends to push the $\phi (^7$Be)/$\phi(^8$B)
ratio to higher values rather than the low one of eq. (12),
\begin{equation}
{\phi (^7{\rm{Be}}) \over \phi(^8 {\rm B})} \sim T_c^{-10}.
\end{equation}
Thus the observations seem difficult to reconcile with plausible
solar model variations: one observable ($\phi(^8$B)) requires a cooler
core while a second, the ratio $\phi(^7$Be)/$\phi(^8$B), requires a hotter one.

An initial question is whether this problem remains significant
when one takes into account known uncertainties in the
parameters of the SSM.
While a detailed summary of standard model uncertainties
would take us well beyond the limits of these lectures, a
qualitative discussion of pp chain nuclear uncertainties is appropriate.
This nuclear microphysics has been the focus of a great deal
of experimental work.  The pp chain involves a series of nonresonant
charged-particle reactions occurring at center-of-mass energies
that are well below the height of the inhibiting Coulomb
barriers.  As the resulting small cross sections preclude
laboratory measurements at the relevant energies, one must
extrapolate higher energy measurements to threshold to
obtain solar cross sections.  This extrapolation is often
discussed in terms of the astrophysical S-factor 
\begin{equation}
\sigma (E) = {S(E) \over E} \exp (-2 \pi \eta)
\end{equation}
where $\eta = {Z_1Z_2 \alpha \over \beta}$, with $\alpha$ the fine structure 
constant and $\beta = v/c$ the relative velocity of the colliding particles.  
This
parameterization removes the gross Coulomb effects associated
with the s-wave interactions of charged, point-like particles.
The remaining energy dependence of S(E) is gentle and
can be expressed as a low-order polynomial in E.  Usually
the variation of S(E) with E is taken from a direct reaction
model and then used to extrapolate higher energy measurements
to threshold.  The model accounts for finite nuclear size
effects, strong interaction effects, contributions from
other partial waves, etc.  As laboratory measurements are
made with atomic nuclei while conditions in the solar core
guarantee the complete ionization of light nuclei, additional
corrections must be made to account for the different
electronic screening environments.

Recently a large working group met at a workshop sponsored by
the Institute for Nuclear Theory, University of Washington, to
review past work on the nuclear reactions of the pp chain
and CNO cycle, to recommend best values and appropriate errors,
and to identify specific issues in experiment and theory
where additional work is needed.
The results will soon be published in Reviews of Modern Physics.
I will not attempt a summary here, but will give one or two
highlights.

The most significant recommend change involves the reaction
$^7$Be(p, $\gamma)\, ^8$B, where the standard S$_{17}$(0)$ \sim$ 22.4 eVb is 
that given~\cite{johnson} by Johnson et al.  Measurements of S$_{17}$(E) are complicated 
by 
the need to use radioactive targets and thus to determine the areal density of 
the $^7$Be target nuclei.  Two techniques have been employed, measuring the 
rate of 478 keV photons from $^7$Be decay or counting the daughter $^7$Li 
nuclei via the reaction $^7$Li (d,p)$^8$Li.  The low-energy data sets 
for 
S$_{17}$(E) disagree by 25\%.  This is a systematic normalization problem as 
each data set is consistent with theory in its dependence on E.  The energy dependence below $\sim$ 500 keV 
is believed to be quite simple as it is determined by the
asymptotic nuclear wave function. 

The Seattle working group on S$_{17}$(E) found that only one low-energy data 
set, that of Filippone et al.~\cite{filippone}, was described in the published literature in 
sufficient detail to be evaluated.  The target activity in that experiment had 
been measured by both 478 keV gamma rays and by the (d,p) reaction, with 
consistent results.  The resulting recommended value was thus based on this 
measurement, yielding
\begin{equation}
S_{17} (0) = 19^{+4}_{-2} \mathrm{eV~b},~~1 \sigma . 
\end{equation}

The $^3$He($\alpha,\gamma) ^7$Be reaction has been measured by two techniques, 
by counting the capture $\gamma$ rays and by detecting the resulting $^7$Be 
activity.  While the two techniques have been used by several groups and have 
yielded separately consistent results, the capture $\gamma$ ray value 
S$_{17}$(0) = 0.507 $\pm $ 0.016 keV b is not in good agreement with the $^7$Be 
activity value 0.572 $\pm$ 0.026 keV-b.  The Seattle working group concluded 
that the evidence for a systematic discrepancy of unknown origin was 
reasonably strong and recommended that standard procedures be used in 
assigning a suitably expanded error.  The recommended value S$_{34}$ (0) is 
0.53 $\pm$ 0.05.

These and other recommended values were recently incorporated 
into the 
BP98 solar model calculation.  While the workshop's
recommended values involve no qualitative changes, there is some broadening of 
error bars.  The downward shift in S$_{17}$(0) leads to a lower 
$^8$B flux.  The workshop's Reviews of Modern Physics article summarizes a 
substantial amount of work on topics not discussed here:  screening effects, 
weak radiative corrections to and exchange current effects on p+p, the atomic 
physics of $^7$Be + e$^-$, etc.  Much of this discussion was useful in 
evaluating possible uncertainties in solar microphysics, and in identifying
opportunities for reducing these uncertainities.

Are uncertainties in the parameters of the SSM a significant
source of uncertainty?  The S-factors discussed above comprise
one set of parameters, but there are others:
the solar lifetime, the opacities, the solar luminosity, etc.
In order to answer this question while also taking into account
correlations among the fluxes when input parameters are
varied, first Bahcall and Ulrich~\cite{bu} and later Bahcall and Haxton~\cite{bh}
constructed 1000 SSMs by
randomly varying five input parameters, the primordial 
heavy-element-to-hydrogen ratio Z/X and S(0) for the  p-p,
$^3$He-$^3$He, $^3$He-$^4$He, and p-$^7$Be reactions, assuming for
each parameter a normal distribution with the mean and standard
deviation.  (These were the parameters
assigned the largest uncertainties.)  Smaller uncertainties
from radiative opacities, the solar luminosity, and the solar
age were folded into the results of the model calculations
perturbatively.

The resulting pattern of $^7$Be and $^8$B flux predictions
is shown in Fig. 3.
The elongated error ellipses indicate that the fluxes are
strongly correlated.  Those variations producing $\phi(^8$B)
below 0.8$\phi^{\rm{SSM}}(^8$B) tend to produce a reduced $\phi(^7$Be), but the
reduction is always less than 0.8.  Thus a greatly reduced 
$\phi(^7$Be) cannot be achieved within the uncertainties assigned
to parameters in the SSM.
    
\begin{figure}[htb]
\psfig{bbllx=0.5cm,bblly=4.0cm,bburx=17cm,bbury=22.5cm,figure=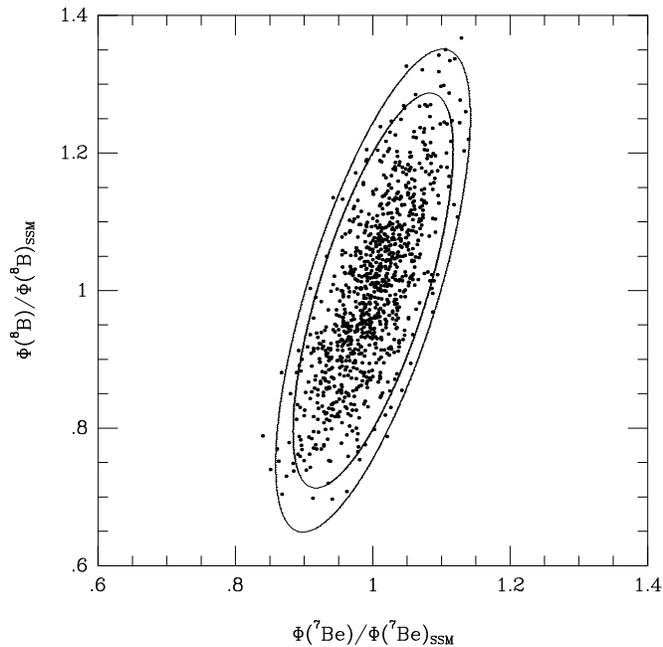,height=3.5in}
\caption{SSM $^7$Be and $^8B$ flux predictions.  The dots represent
the results of SSM calculations where the input parameters were
varied according to their assigned uncertainties, as described
in the text.  The 90\% and 99\% confidence level error ellipses
are shown.}
\end{figure}
  
A similar exploration, but including parameter variations very
far from their preferred values, was carried out by Castellani
et al.~\cite{cast},
who displayed their results as a function of the resulting core temperature
$T_c$.  The pattern that emerges is striking (see Fig. 4):
parameter variations producing the same value of 
$T_c$ produce
remarkably similar fluxes.  Thus 
$T_c$ provides an excellent
one-parameter description of standard model perturbations.
Figure 4 also illustrates the difficulty of producing a 
low ratio of $\phi(^7$Be)/$\phi(^8$B) when 
$T_c$ is reduced.

The 1000-solar-model variations were made under the
constraint of reproducing the solar luminosity.  Those variations show 
a similar strong correlation with $T_c$
\begin{equation}
\phi(pp) \propto T_c^{-1.2} ~~~~~~~  \phi(^7{\rm {Be}}) \propto T_c^8 ~~~~~~~
 \phi(^8 {\rm B}) \propto T_c^{18}.
\end{equation}
Figures 3 and 4 offer a strong argument that reasonable
variations in the parameters of the SSM, or nonstandard
changes in quantities like the metallicity, opacities, or
solar age, cannot produce the pattern of fluxes deduced
from experiment (eq. (12)).  This would seem to limit 
possible solutions to errors either in the underlying physics
of the SSM or in our understanding of neutrino properties. 

\begin{figure}[htb]
\psfig{bbllx=0.3cm,bblly=4.0cm,bburx=14.5cm,bbury=24.0cm,figure=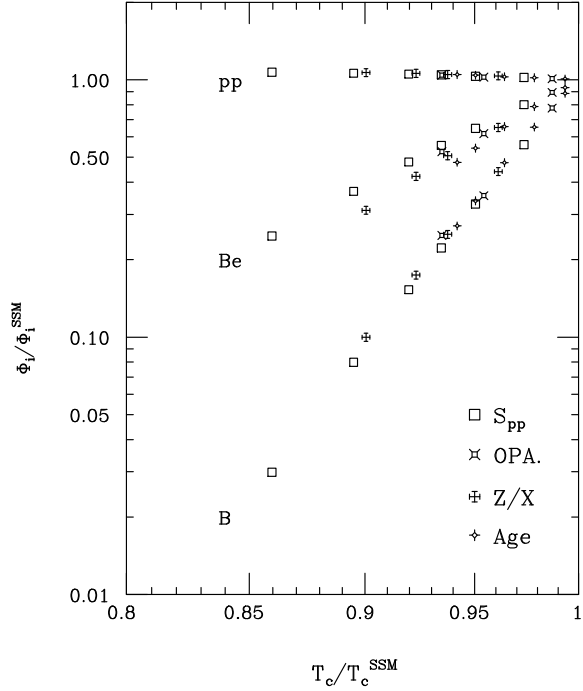,height=3.6in}
\caption{The responses of the pp, $^7$Be, and $^8$B neutrino fluxes
to the indicated variations in solar model input parameters,
displayed as a function of the resulting central temperature
$T_c$.  From Castellani et al.}
\end{figure}
  
\subsection{Nonstandard Solar Models}
Nonstandard solar models include both variations of SSM
parameters far outside the ranges that are generally believed 
to be reasonable (some examples of which are given in Figure 4),
and changes in the underlying physics of the model.  The 
solar neutrino problem has been a major stimulus to models:
in fact, most suggestions were motivated by the hope of
producing a cooler sun ($T_c \sim 0.95 T_c$) that would avoid conflict with the
results of the $^{37}$Cl experiment. 
The suggestions included models with low heavy element
abundances (``low Z" models), in which one abandons the SSM 
assumption that the initial heavy element abundances are those
we measure today at the sun's surface; periodically mixed 
solar cores; models where hydrogen is continually mixed into
the core by turbulent diffusion or by convective 
mixing; and models where the solar core is partially supported by a
strong central magnetic field or by its rapid rotation, 
thereby relaxing the SSM assumption that hydrostatic equilibrium
is achieved only through the gas pressure gradient.  A larger list
is given by Bahcall and Davis~\cite{bd82}.  To illustrate the kinds
of consequences such models have, two of these suggestions
are discussed in more detail below.

In low-Z models one postulates a
reduction in the core metallicity from Z $\sim$ 0.02 to Z $\sim$ 0.002.
This lowers the core opacity (primarily because metals
are very important to free-bound electron transitions),
thus reducing 
$T_c$ and weakening the  ppII and  ppIII cycles.
The attractiveness of low-Z models is due in part to the
existence of mechanisms for adding heavier elements to the
sun's surface.  These include the infall of comets and other
debris, as well as the accumulation of dust as the sun 
passes through interstellar clouds.  However, the increased radiative
energy transport in low-Z models leads to a thin convective
envelope, in contradiction to interpretations of the
5-minute solar surface oscillations.  A low He mass fraction
also results.  As
diffusion of material from a thin convective envelope into
the interior would deplete heavy elements at the surface,
investigators have also questioned whether present abundances could have accumulated 
in low-Z models.  Finally, the general consistency of solar
heavy element abundances with those observed in other main
sequence stars makes the model appear contrived.

Models in which the solar core ($\sim$ 0.2 M$_\odot$) is 
intermittently mixed break the standard model assumption of
a steady-state sun: for a period of several million years
(the thermal relaxation time for the core)
following mixing, the usual relationship between the observed
surface luminosity and rate of energy (and neutrino) production
is altered as the sun burns out of equilibrium.  Calculations
show that both the luminosity and the  $^8$B neutrino flux are
suppressed while the sun relaxes back to the steady state. 
Such models have been considered seriously because of
instabilities associated with large gradients in the $^3$He
abundance, which in equilibrium varies as $\sim T^{-6}$,
where $T$ is the local temperature.
The resulting steep profile is unstable
under finite amplitude displacements of a volume to
smaller r: the energy released by the increased $^3$He burning
at higher T can exceed the energy in the perturbation.
For a discussion of the plausibility of such a trigger for
core mixing, one can see the original work of Dilke and Gough~\cite{gough}
as well as a more recent critique by Merryfield~\cite{merry}.  The
possibility that continuous mixing on time scales of $^3$He
mixing could produce a flux pattern close to that observed
(e.g., a suppression in both the $^8$B neutrino flux and
the $^7$Be/$^8$B flux ratio) was recently discussed by
Cumming and Haxton~\cite{cumming}.
   
This discussion of two of the more seriously explored
nonstandard model possibilities illustrates how changes motivated by the
solar neutrino problem often produce other, unwanted 
consequences.  In particular, many experts feel that the good SSM
agreement with helioseismology is likely to be destroyed by
changes such as those discussed above.

Figure 5 is an illustration by Hata et al.~\cite{hata} of the flux
predictions of several nonstandard models, including a low-Z
model consistent with the $^{37}$Cl results.  As in the Castellani
et al. exploration, the results cluster along a track that
defines the naive 
$T_c$ dependence of the $\phi (^7$Be)/$\phi(^8$B) ratio,
well separated from the experimental contours.
   
\begin{figure}[htb]
\psfig{bbllx=2.3cm,bblly=7.5cm,bburx=17cm,bbury=19.8cm,figure=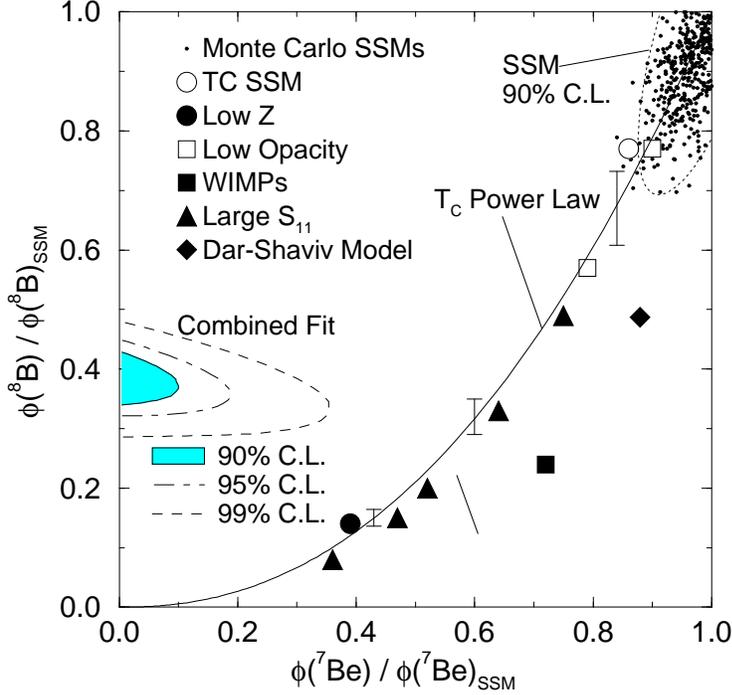,height=3.6in}
\caption{The fluxes allowed by the combined results of the various
solar neutrino experiments compared to the results of SSM 
variations and various nonstandard solar models.  The solid line
in the naive $T_c$ power law discussed in the text.  From
Hata et al.}
\end{figure}
  
There is now a popular argument that no such nonstandard model
can solve the solar neutrino problem: if one assumes undistorted
neutrino spectra, no combination of pp, $^7$Be, and $^8$B
neutrino fluxes fits the experimental results well~\cite{karsten}.  In
fact, in an unconstrained fit, the required $^7$Be flux
is unphysical, negative by almost 3$\sigma$.  Thus, barring
some unfortunate experimental error, it appears we are forced
to look elsewhere for a solution.

If experimental error, SSM parameter uncertainties, and nonstandard
solar physics are ruled out as potential solutions, new particle
physics is left as the leading possibility.
Suggested particle physics solutions of the solar
neutrino problem include neutrino oscillations, neutrino
decay, neutrino magnetic moments, and weakly interacting
massive particles.  Among these, the Mikheyev-Smirnov-Wolfenstein 
effect --- neutrino oscillations enhanced by matter
interactions --- is widely regarded as the most plausible.

\subsection{Neutrino Oscillations}
One odd feature of particle physics is that neutrinos,
which are not required by any symmetry to be massless, nevertheless
must be much lighter than any of the other known fermions.
For instance, the current limit on the $\overline{\nu}_e$ mass is $\lsim$ 5 eV.
The standard model requires neutrinos to be massless, but the
reasons are not fundamental.  Dirac mass terms $m_D$, analogous
to the mass terms for other fermions, cannot be constructed
because the model contains no right-handed neutrino fields.
Neutrinos can also have Majorana mass terms
\begin{equation}
\overline{\nu^c_L} m_L \nu_L ~~~ \mathrm{and} ~~~ \overline{\nu^c_R} m_R \nu_R 
\end{equation}
where the subscripts $L$ and $R$ denote left- and right-handed projections
of the neutrino field $\nu$, and the superscript $c$ denotes charge conjugation.
The first term above is constructed from left-handed fields, but can 
only arise as a nonrenormalizable effective interaction when
one is constrained to generate $m_L$ with the doublet scalar field of
the standard model.  The second term is absent from the standard  
model because there are no right-handed neutrino fields.

None of these standard model arguments 
carries over to the more general, unified theories that 
theorists believe will supplant the standard model.  
In the enlarged multiplets of extended models it is
natural to characterize the fermions of a single family,
e.g., $\nu_e$, e, u, d, by the same mass scale $m_D$.  Small neutrino
masses are then frequently explained as a result of the
Majorana neutrino masses.  In the seesaw mechanism,
\begin{equation}
M_\nu \sim \left(\begin{array}{cc}
0 & m_D \\
m^T_D & m_R \end{array}\right) .  
\end{equation}
Diagonalization of the mass matrix  
produces one light neutrino, $m_{\mathrm{light}}\sim {m_D^2 \over m_R}$, and one 
unobservably
heavy, $m_{\mathrm{heavy}} \sim m_R$.  The factor ($m_D$/$m_R$) is the needed small
parameter that accounts for the distinct scale of neutrino
masses.  The masses for the $\nu_e, \nu_\mu$, and $\nu_\tau$ are then
related to the squares of the corresponding quark masses
$m_u$, $m_c$, and $m_t$.  Taking $m_R \sim 10^{16}$ GeV, a typical grand 
unification scale for models built on groups like SO(10), the seesaw mechanism gives the crude relation
\begin{equation}
m_{\nu_e}: m_{\nu_\mu}: m_{\nu_\tau} \leftrightarrow 2 \cdot 10^{-12}: 2 
\cdot 10^{-7}: 3 \cdot 10^{-3} \mathrm{eV}. 
\end{equation}
The fact that solar neutrino experiments can probe small
neutrino masses, and thus provide insight into possible new
mass scales $m_R$ that are far beyond the reach of direct
accelerator measurements, has been an important theme of
the field.
    
Now one of the most interesting possibilities for solving the
solar neutrino problem has to do with neutrino masses.  For
simplicity we will discuss just two neutrinos.  If a neutrino
has a mass $m$, we mean that as it propagates through free 
space, its energy and momentum are related in the usual way
for this mass.  Thus if we have two neutrinos, we can label 
those neutrinos according to the eigenstates of the free 
Hamiltonian, that is, as mass eigenstates.

But neutrinos are produced by the weak interaction.  In this 
case, we have another set of eigenstates, the flavor eigenstates.
We can define a $\nu_e$ as the neutrino that accompanies the positron in
$\beta$ decay.  Likewise we label by $\nu_\mu$ the neutrino 
produced in muon decay.

Now the question: are the eigenstates of the free Hamiltonian 
and of the weak interaction Hamiltonian identical?  Most likely
the answer is no: we know this is the case with the quarks,
since the different families (the analog of the mass eigenstates)
do interact through the weak interaction.  That is, the up
quark decays not only to the down quark, but also occasionally
to the strange quark.  (This is why we had a $\cos \theta_c$
in our weak interaction amplitude: the amplitude for $u \rightarrow s$
is proportional to $\sin \theta_c$.)  Thus we suspect that 
the weak interaction and mass eigenstates, while spanning the
same two-neutrino space, are not coincident:
the mass eigenstates $|\nu_1 \rangle$ and $|\nu_2 \rangle$ (with masses
$m_1$ and $m_2$) are related to the weak interaction eigenstates by
\begin{eqnarray}
|\nu_e\rangle  &=& \cos \theta_v |\nu_1\rangle  + \sin \theta_v|\nu_2 \rangle  \nonumber \\
|\nu_\mu\rangle &=& - \sin \theta_v |\nu_1 \rangle + \cos \theta_v |\nu_2 
\rangle 
\end{eqnarray}
where $\theta_v$ is the (vacuum) mixing angle. 
  
An immediate consequence is that a state produced as a $|\nu_e\rangle$
or a $|\nu_\mu\rangle$ at some time $t$ --- for example, a neutrino
produced in $\beta$ decay --- does not remain a pure flavor eigenstate
as it propagates away from the source.  This is because the different
mass eigenstates comprising the neutrino will accumulate different
phases as they propagate downstream, a phenomenon known as 
vacuum oscillations (vacuum because the experiment is done in free
space).  To see the effect, suppose we produce a neutrino in some
$\beta$ decay where we measure the momentum of the initial nucleus,
final nucleus, and positron.  Thus the outgoing neutrino is a
momentum eigenstate~\cite{nauenberg}.  At time $t$=0
\begin{equation}
|\nu(t=0)\rangle  = |\nu_e \rangle = \cos \theta_v |\nu_1\rangle  + \sin \theta_v|\nu_2 \rangle . 
\end{equation}
Each eigenstate subsequently propagates with a phase
\begin{equation}
e^{i(\vec{k} \cdot \vec{x} - \omega t)} =
e^{i(\vec{k} \cdot \vec{x} - \sqrt{m_i^2 + k^2}t)} . 
\end{equation}
But if the neutrino mass is small compared to the neutrino 
momentum/energy, one can write
\begin{equation}
\sqrt{m_i^2+k^2} \sim k(1 + {m_i^2 \over 2k^2}) . 
\end{equation}
Thus we conclude
\begin{eqnarray}
|\nu(t) \rangle &=& e^{i(\vec{k} \cdot \vec{x} - kt
-(m_1^2+m_2^2)t/4k)} \nonumber \\
& & \times [\cos \theta_v |\nu_1 \rangle e^{i \delta m^2 t/4k}
+ \sin \theta_v |\nu_2 \rangle e^{-i \delta m^2 t/4k} ] . 
\end{eqnarray}
We see there is a common average phase (which has no physical
consequence) as well as a beat phase that depends on
\begin{equation}
\delta m^2 = m_2^2 - m_1^2 .
\end{equation}
Now it is a simple matter to calculate the probability that 
our neutrino state remains a $|\nu_e\rangle$ at time t
\begin{eqnarray}
P_{\nu_e} (t) &=& | \langle \nu_e | \nu(t) \rangle |^2 \nonumber \\ 
 &=& 1 - \sin^2 2 \theta_v \sin^2 \left({\delta m^2 t \over 4 
k}\right) \rightarrow 1 - {1 \over 2} \sin^2 2 \theta_v 
\end{eqnarray}
where the limit on the right is appropriate for large $t$.
Now $E \sim k$, where E is the neutrino energy, by our assumption
that the neutrino masses are small compared to $k$.  Thus we can
reinsert the units above to write the probability in terms of
the distance $x$ of the neutrino from its source,
\begin{equation}
P_{\nu} (x) =1 - \sin^2 2 \theta_v \sin^2 \left({\delta m^2c^4 x\over 4 
\hbar c E} \right) . 
\end{equation}
(When one properly describes the neutrino state as a wave packet,
the large-distance behavior follows from the eventual separation
of the mass eigenstates.)  If the
the oscillation length
\begin{equation}
L_o = {4 \pi \hbar c E \over \delta m^2 c^4} 
\end{equation}
is comparable to or shorter than one astronomical unit, a 
reduction in the solar $\nu_e$ flux would be expected in terrestrial
neutrino oscillations.
  
The suggestion that the solar neutrino problem could
be explained by neutrino oscillations was first made by
Pontecorvo in 1958, who pointed out the analogy with $K_0 \leftrightarrow \bar 
K_0$     
oscillations.  From the point of view of particle physics, 
the sun is a marvelous neutrino source.  The neutrinos travel a long
distance and have low energies ($\sim$ 1 MeV), implying a sensitivity to
\begin{equation}
\delta m^2 \gsim 10^{-12} eV^2.
\end{equation}
In the seesaw mechanism, $\delta m^2 \sim m^2_2$, so neutrino masses as
low as $m_2 \sim 10^{-6}$ eV could be probed.  In contrast, terrestrial
oscillation experiments with accelerator or reactor
neutrinos are typically limited to $\delta m^2 \gsim 0.1 $ eV$^2$. 

From the expressions above one expects vacuum oscillations to affect
all neutrino species equally, if the oscillation length is small
compared to an astronomical unit.  This is somewhat in conflict
with the data, as we have argued that the $^7$Be neutrino flux
is quite suppressed.
Furthermore, there is a weak theoretical prejudice that $\theta_v$ should be
small, like the Cabibbo angle.
The first objection, however, can be circumvented in
the case of ``just so" oscillations where the oscillation 
length is comparable to one astronomical unit.
In this case the oscillation probability becomes sharply
energy dependent, and one can choose $\delta m^2$ to preferentially
suppress one component (e.g., the monochromatic $^7$Be neutrinos).
This scenario has been explored by several groups and
remains an interesting possibility.  However, the
requirement of large mixing angles remains.
  
Below we will see that stars allow us to ``get around" this 
problem with small mixing angles.  In preparation for this, we
first present the results above in a slightly more general
way.  The analog of eq. (24) for an initial
muon neutrino ($|\nu(t=0)\rangle = |\nu_\mu\rangle$) is
\begin{eqnarray}
|\nu(t) \rangle &=& e^{i(\vec{k} \cdot \vec{x} - kt
-(m_1^2+m_2^2)t/4k)} \nonumber \\
&& \times [-\sin \theta_v |\nu_1 \rangle e^{i \delta m^2 t/4k}
+ \cos \theta_v |\nu_2 \rangle e^{-i \delta m^2 t/4k} ]
\end{eqnarray}
Now if we compare eqs. (24) and (30) we see that they are special cases
of a more general problem.  Suppose we write our initial neutrino
wave function as
\begin{equation}
 |\nu(t=0)\rangle = a_e(t=0) |\nu_e \rangle + a_\mu(t=0) 
|\nu_\mu \rangle . 
\end{equation}
Then eqs. (24) and (30) tell us that the subsequent propagation is described
by changes in $a_e(x)$ and $a_\mu(x)$ according to
(this takes a bit of algebra)
\begin{equation}
i {d \over dx} \left( \matrix { a_{\textstyle e} \cr
a_{\textstyle \mu} \cr} \right) = {1 \over 4E} \left ( \matrix{
- \delta m^2 \cos 2 \theta_{\textstyle v}
~~~~~~~~~~~\delta m^2\sin
2\theta_{\textstyle v} \cr 
\delta m^2\sin 2 \theta_{\textstyle v} ~~~~~~~~~~~ 
\delta m^2
\cos 2\theta_{\textstyle v} \cr} \right) \left( \matrix {
a_{\textstyle e} \cr
a_{\textstyle \mu} \cr} \right) . 
\end{equation}
Note that the common phase has been ignored: it can be absorbed
into the overall phase of the coeeficients $a_e$ and $a_\mu$,
and thus has no consequence.   
Also, we have equated $x = t,$ that is, set $c$ = 1.

\subsection{The Mikheyev-Smirnov-Wolfenstein Mechanism}
The view of neutrino oscillations changed 
when Mikheyev and Smirnov~\cite{ms} showed in 1985 that the
density dependence of the neutrino effective mass, a phenomenon
first discussed by Wolfenstein in 1978, could greatly enhance
oscillation probabilities: a $\nu_e$ is adiabatically transformed
into a $\nu_\mu$ as it traverses a critical density within the sun.
It became clear that the sun was not only an excellent 
neutrino source, but also a natural regenerator for cleverly
enhancing the effects of flavor mixing. 
   
While the original work of Mikheyev and Smirnov was
numerical, their phenomenon was soon understood analytically
as a level-crossing problem.  If one writes the neutrino
wave function in matter as in eq. (31),
the evolution of $a_e(x)$ and $a_\mu(x)$ is governed by
\begin{equation}
i {d \over dx} \left( \matrix { a_{\textstyle e} \cr
a_{\textstyle \mu} \cr} \right) = {1 \over 4E} \left ( \matrix{
2E \sqrt2 G_F \rho(x) - \delta m^2 \cos 2 \theta_{\textstyle v}
~~~~~~\delta m^2\sin
2\theta_{\textstyle v} \cr 
\delta m^2\sin 2 \theta_{\textstyle v} ~~~ -2E \sqrt2 G_F \rho(x) +
\delta m^2
\cos 2\theta_{\textstyle v} \cr} \right) \left( \matrix {
a_{\textstyle e} \cr
a_{\textstyle \mu} \cr} \right) 
\end{equation}
where G$_F$ is the weak coupling constant and $\rho (x)$ the solar
electron density.  If $\rho (x)$ = 0, this is exactly our previous
result and can be trivially
integrated to give the vacuum oscillation solutions of Sec. 2.5.
The new contribution to the diagonal elements, $2 E \sqrt2 G_F \rho(x)$, 
represents the effective contribution to $M^2_\nu$  that arises 
from neutrino-electron scattering.  The indices of refraction
of electron and muon neutrinos differ because the former
scatter by charged and neutral currents, while the latter 
have only neutral current interactions.  The difference in
the forward scattering amplitudes determines the density-dependent
splitting of the diagonal elements of the new matter equation. 

It is helpful to rewrite this equation in a basis consisting of the light and heavy 
local mass eigenstates (i.e., the states that diagonalize the right-hand side 
of the equation),
\begin{eqnarray}
|\nu_L (x)\rangle &=& \cos \theta (x)|\nu_e\rangle - \sin \theta 
(x)|\nu_\mu\rangle \nonumber \\
|\nu_H(x)\rangle &=& \sin \theta (x)|\nu_e\rangle + \cos \theta (x)|\nu_\mu 
\rangle . 
\end{eqnarray}
The local mixing angle is defined by
\begin{eqnarray}
\sin 2 \theta (x)  &=& {\sin 2 \theta_{\textstyle v} \over \sqrt{X^2 (x) + \sin^2
2\theta_{\textstyle v}}} \nonumber \\
\cos 2\theta (x)  &=& {-X (x) \over \sqrt{X^2 (x) + \sin^2 2\theta_{\textstyle
v}}} 
\end{eqnarray}
where $X(x) = 2 \sqrt2G_F \rho(x) E/\delta m^2 - \cos 2\theta_{\textstyle v}$.
Thus
$\theta(x)$ ranges from $\theta_{\textstyle v}$ to $\pi/2$ as the density
$\rho(x)$ goes
from 0 to $\infty$. 

If we define
\begin{equation}
|\nu (x) \rangle = a_H(x)|\nu_H(x)\rangle + a_L(x)|\nu_L(x)\rangle,
\end{equation}
the neutrino propagation can be rewritten in terms of the local
mass eigenstates
\begin{equation}
i {d \over dx} \pmatrix{
a_H \cr
a_L \cr} = \pmatrix {
\lambda(x) & i \alpha (x) \cr
-i \alpha (x) & - \lambda (x) \cr }
\pmatrix
{a_H \cr
a_L }
\end{equation}
with the splitting of the local mass eigenstates determined by
\begin{equation}
2 \lambda (x) = {\delta m^2 \over 2E} \sqrt{X^2 (x) + \sin^2 2 
\theta_{\textstyle v}} 
\end{equation}
and with mixing of these eigenstates governed by the density gradient
\begin{equation}
\alpha (x) = \left({E \over \delta m^2}\right)
 \, {\sqrt2 \, G_F {d \over dx}
\rho(x)
\sin 2 \theta_{\textstyle v} \over X^2 (x) + \sin^2 2 \theta_{\textstyle v}}.
\end{equation}
The results above are quite interesting: the local mass eigenstates
diagonalize the matrix if the density is constant.  In such a limit,
the problem is no more complicated than our original vacuum
oscillation case, although our mixing angle is changed because of
the matter effects.  But if the density is not constant, the
mass eigenstates in fact evolve as the density changes.  This
is the crux of the MSW effect.
Note that the splitting achieves
its minimum value, ${\delta m^2 \over 2E} \sin 2 \theta_v$, at a critical density $\rho_c =
\rho (x_c)$
\begin{equation}
2 \sqrt2 E G_F \rho_c = \delta m^2 \cos 2 \theta_v 
\end{equation}
that defines the point where the diagonal elements of the original flavor matrix cross. 

Our local-mass-eigenstate form of the propagation equation can be trivially integrated if the splitting of the diagonal
elements is 
large compared to the off-diagonal elements,
\begin{equation}
\gamma (x) = \left|{\lambda (x) \over \alpha (x)}\right| = {\sin^2
2\theta_{\textstyle v} \over \cos
2\theta_{\textstyle v}} \, {\delta m^2 \over 2 E} \, {1 \over |{1 \over \rho_c}
{d \rho (x) \over
dx}|} {[X (x)^2 + \sin^2 2\theta_v]^{3/2} \over \sin^3 2\theta_v} \gg 1, 
\end{equation}
a condition that becomes particularly stringent near the crossing point,
\begin{equation}
\gamma_c = \gamma (x_c) = {\sin^2 2\theta_v \over \cos 2\theta_v} \, {\delta
m^2 \over 2 E} \, {1 \over \left|{1 \over \rho_c} {d \rho (x) \over dx}|_{x =
x_c}\right|} 
\gg 1. 
\end{equation}
The resulting adiabatic electron neutrino survival probability~\cite{bethe}, valid when
$\gamma_c \gg 1$, is
\begin{equation}
P^{\rm adiab}_{\nu_e} = {1 \over 2} + {1 \over 2} \cos 2 \theta_v \cos 2
\theta_i 
\end{equation}
where $\theta_i = \theta (x_i)$ is the local mixing angle at the density where
the neutrino was produced. 

\begin{figure}[htb]
\psfig{bbllx=1.2cm,bblly=2.0cm,bburx=18cm,bbury=14.5cm,figure=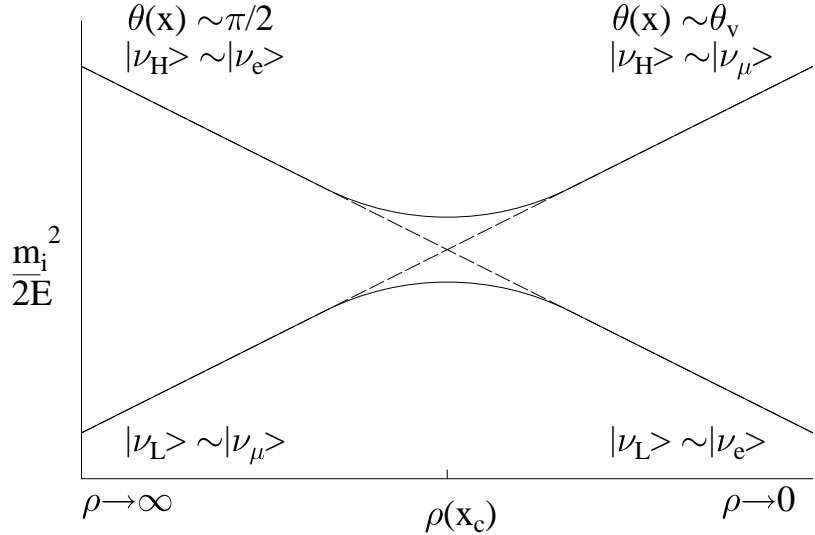,height=3.0in}
\caption{Schematic illustration of the MSW crossing.  The dashed 
lines correspond to the electron-electron and muon-muon diagonal
elements of the $M^2$ matrix in the flavor basis.  Their 
intersection defines the level-crossing density $\rho_c$.
The solid lines are the trajectories of the light and heavy
local mass eigenstates.  If the electron neutrino is produced 
at high density and propagates adiabatically, it will follow
the heavy-mass trajectory, emerging from the sun as a $\nu_\mu$.}
\end{figure}
  
The physical picture behind this derivation is illustrated
in Fig. 6.  One makes the usual assumption that, in vacuum,
the $\nu_e$ is almost identical to the light mass eigenstate,
$\nu_L(0)$, i.e., $m_1 < m_2$ and $\cos \theta_v \sim$ 1.  But as the density increases,
the matter effects make the $\nu_e$ heavier than the $\nu_\mu$, with $\nu_e 
\to \nu_H (x)$  as $\rho(x)$ becomes large.  The special property of 
the sun is that it produces $\nu_e$s at high density that then propagate to 
the vacuum where they
are measured.  The adiabatic approximation tells us that if
initially $\nu_e \sim \nu_H (x)$, the neutrino will remain on the heavy
mass trajectory provided the density changes slowly.
That is, if the solar density gradient is sufficiently gentle,
the neutrino will emerge from the sun as the heavy vacuum
eigenstate, $ \sim \nu_\mu$.  This guarantees nearly complete conversion
of $\nu_e$s into $\nu_\mu$s, producing a flux that cannot be detected
by the Homestake or SAGE/GALLEX detectors. 
   
But this does not explain the curious pattern of partial
flux suppressions coming from the various solar neutrino experiments.  The key to this is the behavior when
$\gamma_c \lsim$ 1.  Our expression for $\gamma(x)$ shows that the critical region
for nonadiabatic behavior occurs in a narrow region (for small $\theta_v$)
surrounding the crossing point, and that this behavior is 
controlled by the derivative of the density.  This suggests an
analytic strategy for handling nonadiabatic crossings: one
can replace the true solar density by a simpler (integrable!) two-parameter 
form that is constrained to reproduce the true density and its derivative at 
the crossing point $x_c$. Two convenient choices are the linear $(\rho (x) = a 
+ bx)$ and exponential $(\rho (x) = ae^{-bx})$ profiles.  As the density 
derivative at $x_c$ governs the nonadiabatic behavior, this procedure should 
provide an accurate description of the hopping probability between the local 
mass eigenstates when the neutrino traverses the crossing point.  The initial 
and ending points $x_i$ and $x_f$ for the artificial profile are then chosen 
so that $\rho(x_i)$ is the density where the neutrino was produced in the 
solar core and $\rho(x_f) = 0$ (the solar surface), as illustrated in in Fig. 7. 
Since the adiabatic result ($P_{\nu_e}^{\mathrm{adiab}}$) depends only on the local mixing angles 
at these points, this choice builds in that limit.  But our original flavor-basis equation can then be integrated 
exactly for linear and exponential profiles, with the results given in terms 
of parabolic cylinder and Whittaker functions, respectively.   

\begin{figure}[htb]
\psfig{bbllx=0.0cm,bblly=2.8cm,bburx=16cm,bbury=21.3cm,figure=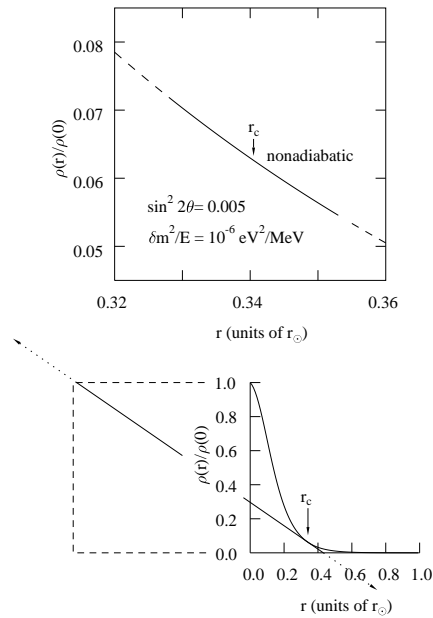,height=3.3in}
\caption{The top figure illustrates, for one choice of sin$^2 2\theta$
and $\delta m^2$, that the region of nonadiabatic propagation
(solid line) is usually confined to a narrow region around the
crossing point $r_c$.  In the lower figure, the solid lines
represent the solar density and a linear approximation to that 
density that has the correct initial and final values, as
well as the correct density and density derivative at $r_c$.
Thus the linear profile is a very good approximation to the
sun in the vicinity of the crossing point.  The MSW equations 
can be solved analytically for this wedge.  By extending the
wedge to $\pm \infty$ (dotted lines) and assuming adiabatic
propagation in these regions of unphysical density, one obtains
the simple Landau-Zener result discussed in the text.}
\end{figure}
  
That result can be simplified further
by observing that the nonadiabatic region is generally confined to 
a narrow region around $x_c$, away from the endpoints $x_i$ and $x_f$.  We  
can then extend the artificial profile to $x = \pm \infty$, as illustrated by  
the dashed lines in Fig. 7.  As the neutrino propagates adiabatically in the 
unphysical region $x < x_i$, the exact soluation in the physical region can be 
recovered by choosing the initial boundary conditions
\begin{eqnarray}
a_L (- \infty) &=& - a_\mu (- \infty) = \cos \theta_i e^{- i \int^{x_i}_{- 
\infty} \lambda (x) dx} \nonumber\\
a_H (- \infty) &=& a_e (- \infty) = \sin \theta_i e^{i \int^{x_i}_{- \infty} 
\lambda (x) dx} . 
\end{eqnarray}
That is, $|\nu (-\infty)\rangle$ will then adiabatically evolve to $|\nu 
(x_i)\rangle = |\nu_e\rangle$ as $x$ goes from $- \infty$ to $x_i$.  The 
unphysical region $x > x_f$ can be handled similarly.  

With some algebra a simple generalization of the adiabatic
result emerges that is valid for all $\delta m^2/E$ and $\theta_v$
\begin{equation}
P_{\nu_e} = {1 \over 2} + {1 \over 2} \cos 2 \theta_v \cos 2 \theta_i ( 1 - 
2P_{\rm {hop}}) 
\end{equation}
where P$_{\rm {hop}}$ is the Landau-Zener probability of hopping from the heavy mass
trajectory to the light trajectory on traversing the crossing
point.  For the linear approximation to the density~\cite{hlz,plz},
\begin{equation}
P^{\rm {lin}}_{\rm {hop}} = e^{- \pi \gamma_c/2} . 
\end{equation}
As it must by our construction, $P_{\nu_e}$ reduces to P$^{\rm 
{adiab}}_{\nu_e}$ for $\gamma_c \gg$ 1.   
When the crossing becomes nonadiabatic (e.g., $\gamma_c \ll 1$ ),
the hopping probability goes to 1, allowing the neutrino to
exit the sun on the light mass trajectory as a $\nu_e$, i.e., no conversion 
occurs. 

Thus there are two conditions for strong           
conversion of solar neutrinos:  there must be a level 
crossing (that is, the solar core density must be sufficient 
to render $\nu_e \sim \nu_H (x_i)$  when it is first
produced) and the crossing must be adiabatic.  The first
condition requires that $\delta m^2/E$ not be too large, and the 
second $\gamma_c \gsim$ 1.  The combination of these two constraints,
illustrated in Fig. 8, defines a triangle of interesting
parameters in the ${\delta m^2 \over E} - \sin^2 2\theta_v$ plane, as Mikheyev and Smirnov
found by numerically
integration.  A remarkable feature of this triangle
is that strong $\nu_e \to \nu_\mu$ conversion can occur for very small
mixing angles $(\sin^2 2 \theta \sim10^{-3}$), unlike the vacuum case. 

\begin{figure}[htb]
\psfig{bbllx=-1.5cm,bblly=0.0cm,bburx=15cm,bbury=22.0cm,figure=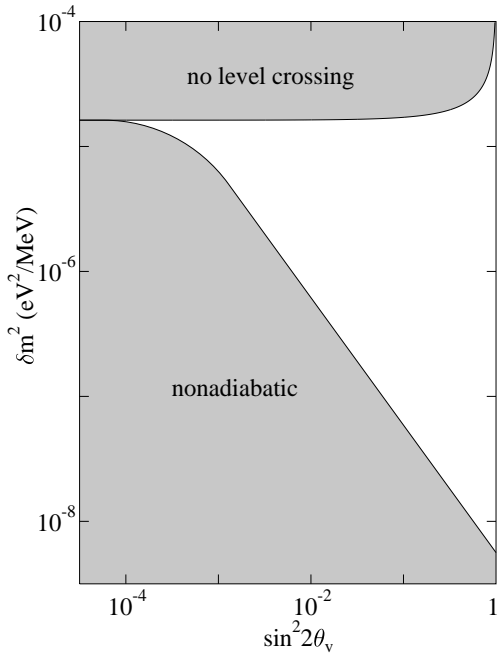,height=3.6in}
\caption{MSW conversion for a neutrino produced at the sun's
center.  The upper shaded region indices thoses $\delta m^2/E$
where the vacuum mass splitting is too great to be overcome
by the solar density.  Thus no level crossing occurs.  The
lower shaded region defines the region where the level crossing
is nonadiabatic ($\gamma_c$ less than unity).  The unshaded
region corresponds to adiabatic level crossings where strong
$\nu_e \rightarrow \nu_\mu$ will occur.}
\end{figure}
  
One can envision superimposing on Fig. 8 the spectrum of solar neutrinos, plotted as a 
function of ${\delta m^2 \over E}$ for some choice of $\delta m^2$.
Since Davis sees {\it some} solar neutrinos, the solutions must 
correspond to the boundaries of the triangle in Fig. 8.  The horizontal 
boundary indicates the maximum ${\delta m^2 \over E}$ for which the sun's 
central density is sufficient to cause a level crossing.  If a spectrum 
properly straddles this boundary, we obtain a result consistent with the 
Homestake experiment in which low energy neutrinos (large 1/E) lie above the 
level-crossing boundary (and thus remain $\nu_e$'s), but the high-energy 
neutrinos (small 1/E) fall within the unshaded region where strong conversion 
takes place.  Thus such a solution would mimic nonstandard solar models in 
that only the $^8$B neutrino flux would be strongly suppressed.  The diagonal 
boundary separates the adiabatic and nonadiabatic regions.  If the spectrum 
straddles this boundary, we obtain a second solution in which low energy 
neutrinos lie within the conversion region, but the high-energy neutrinos 
(small 1/E) lie below the conversion region and are characterized by $\gamma 
\ll 1$ at the crossing density.  (Of course, the boundary is not a sharp one, 
but is characterized by the Landau-Zener exponential).  Such a nonadiabatic 
solution is quite distinctive since the flux of  pp neutrinos, which is 
strongly constrained in the standard solar model and in any steady-state 
nonstandard model by the solar luminosity, would now be sharply reduced.  
Finally, one can imagine ``hybrid" solutions where the spectrum straddles both
the level-crossing (horizontal)
boundary and the adiabaticity (diagonal) boundary for small $\theta$,
thereby reducing the $^7$Be neutrino flux more than either the
pp or $^8$B fluxes. 

What are the results of a careful search for MSW solutions
satisfying the Homestake, Kamiokande/SuperKamiokande, and SAGE/GALLEX constraints?
This has been done by several groups: recent results will be
discussed by N. Hata in his lectures.  One
solution, corresponding to a region surrounding $\delta m^2 \sim 6 \cdot 10^{-6} $eV$^2$ 
and $\sin^2 2\theta_v \sim 6 \cdot 10^{-3}$, is the hybrid case described above.  It is commonly 
called the small-angle solution.  A second, large-angle solution
exists, corresponding to $\delta m^2 \sim 10^{-5} $eV$^2$ and $\sin^2 2 \theta_v 
\sim$ 0.6. 
These solutions can be distinguished by their characteristic
distortions of the solar neutrino spectrum.  The survival
probabilities $P_{\nu_e}^{\rm MSW}$(E) for the small- and large-angle parameters
given above are shown as a function of E in Fig. 9.

\begin{figure}[htb]
\psfig{bbllx=0.5cm,bblly=1.3cm,bburx=18cm,bbury=13.7cm,figure=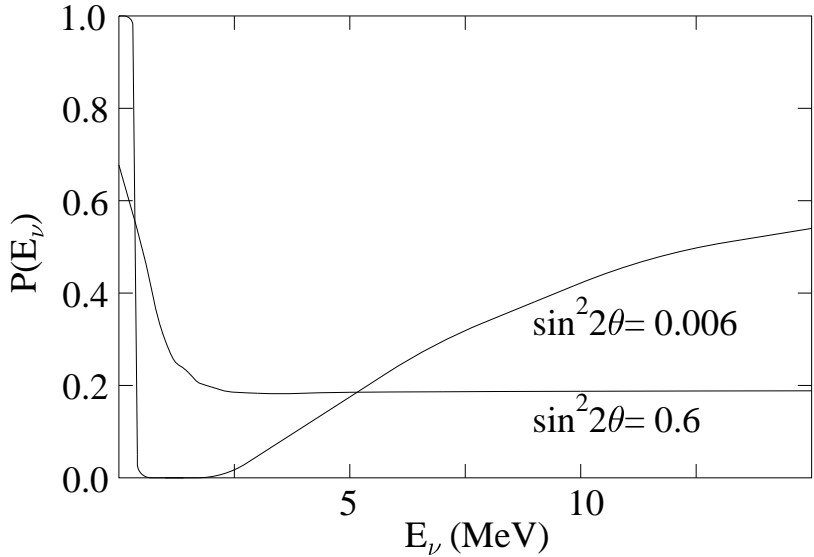,height=3.0in}
\caption{MSW survival probabilities P(E$_\nu$) for typical small
angle and large angle solutions.}
\end{figure}
  
The MSW mechanism provides a natural explanation for the
pattern of observed solar neutrino fluxes.  While it requires
profound new physics, both massive neutrinos and neutrino mixing
are expected in extended models.  The small-angle solution 
corresponds to $\delta m^2 \sim 10^{-5}$ eV$^2$, and thus is consistent with
$m_2 \sim$ few $\cdot 10^{-3}$ eV.  This is a typical $\nu_\tau$ mass in models
where $m_R \sim m_{\rm {GUT}}$.  This mass is also reasonably close to atmospheric neutrino values.  On the other hand, if it is the $\nu_\mu$  
participating in the oscillation, this gives $m_R \sim 10^{12}$ GeV
and predicts a heavy $\nu_\tau \sim$ 10 eV.  Such
a mass is of great interest cosmologically as it would have
consequences for supernova physics,
the dark matter problem, and the formation of large-scale structure. 

\subsection{Spin-Flavor Oscillations}
If the MSW mechanism proves not to be the solution of the solar
neutrino problem, it still will have greatly enhanced the
importance of solar neutrino physics: the existing experiments
have ruled out large regions in the $\delta m^2 - \sin^2 2\theta_v$ plane
(corresponding to nearly complete $\nu_e \to \nu_\mu$ conversion) that
remain hopelessly beyond the reach of accelerator neutrino
oscillation experiments. 
 
A number of other particle physics solutions have been 
considered, such as neutrino decay, matter-catalyzed neutrino
decay, and solar energy transport by weakly interacting massive particles.
But perhaps the most interesting possibility, apart from the
MSW mechanism, was stimulated by suggestions that the $^{37}$Cl
signal might be varying with a period comparable to the 11-year
solar cycle.  While the evidence for this has weakened, 
the original claims generated renewed interest in neutrino
magnetic moment interactions with the solar magnetic field. 

The original suggestions by Cisneros and by
Okun, Voloshyn, and Vysotsky envisioned the rotation
\begin{equation}
\nu_{e_L} \to \nu_{e_R} 
\end{equation}
producing a right-handed neutrino with sterile interactions in
the standard model.
With the discovery of the MSW mechanism, it was realized
that matter effects would break the vacuum degeneracy of the $\nu_{e_L}$
and $\nu_{e_R}$, suppressing the spin precession.  Lim and Marciano~\cite{lim}
and Akmedov~\cite{ak}
pointed out that this difficulty was naturally circumvented
for
\begin{equation}
\nu_{e_L} \to \nu_{\mu_R} 
\end{equation}
as the different matter interactions of the $\nu_e$ and $\nu_\mu$ can
compensate for the vacuum $\nu_e - \nu_\mu$  mass difference, producing a
crossing similar to the usual MSW mechanism.  Such spin-flavor
precession can then occur at full strength due to an off-diagonal
(in flavor) magnetic moment.

Quite relevant to this suggestion is the very strong limit
on both diagonal and off-diagonal magnetic moments is imposed
by studies of the red giant cooling process of plasmon decay
into neutrinos
\begin{equation}
\gamma^* \to \nu_i \bar \nu_j. 
\end{equation}
The result is $|\mu_{ij}| \lsim 3 \cdot 10^{-12} \mu_B$, where $\mu_B$
is an
electron Bohr magneton~\cite{raff}.  (This can be compared to  
a simple one-loop estimate~\cite{fuj} of the neutrino magnetic
moment of $\sim 10^{-18} \mu_B$, 
taking a typical dark matter value for the neutrino mass of a few eV.)
If a magnetic moment at the red giant limit is assumed, it follows that
solar magnetic field strengths of $B_\odot \gsim 10^6 G$ are needed to
produce interesting effects.  Since the location of the 
spin-flavor level crossing depends on the neutrino energy,
such fields have to be extensive to affect an appreciable
fraction of the neutrino spectrum.  It is unclear whether 
these conditions can occur in the sun.
This constraint from stellar cooling leads us naturally into
our next lecture. 
    
\section{Dirac and Majorana Neutrinos and Stellar Cooling}
\subsection{The Neutrino Mass Matrix}
Consider a general 4n $\times$ 4n neutrino mass matrix where 
n is the number of flavors
\begin{equation}
\begin{array}{c} (\bar{\Psi}^c_L,\bar{\Psi}^R,
\bar{\Psi}_L,\bar{\Psi}^c_R) \\ \\ \\ \end{array}
\left( \begin{array}{cccc} 0 & 0 &  M_L & M^T_D \\
0 & 0 & M_D &  M_R^\dag \\  M_L^\dag & M_D^\dag & 0 & 0 \\ M_D^* &  M_R & 0 & 0
\end{array} \right) 
\left( \begin{array}{c} \Psi^c_L \\ \Psi_R \\ \Psi_L \\ \Psi^c_R
\end{array} \right)
\end{equation}
where each entry in this matrix is understood to be a n $\times$ n
matrix operating in flavor space.  The entries $M_D$ are the
Dirac mass terms, while the $M_L$ and $M_R$ are the Majorana terms.
The latter break the invariance of the Dirac equation under the transformation $\psi(x) \rightarrow
e^{i\alpha} \psi(x)$ associated with a conserved lepton number.
Thus it is these terms that govern lepton-number-violating 
processes like double beta decay.

One can proceed to diagonalize this matrix
\begin{equation}
\Psi_{\nu(e)}^L = \sum_{i=1}^{2n} U_{ei}^L \tilde{\nu}_i(x) 
~~\mathrm{with~masses}~m_i .
\end{equation}
The eigenstates are two-component Majorana neutrinos~\cite{haxtonbb},
yielding the proper $2 \times 2n = 4n$ degrees of freedom,
where $n$ is the number of flavors.
We can recover the Majorana and Dirac limits:\\
$\bullet$ If $M_R$ = $M_L$ = 0, the eigenstates of this matrix
become pairwise degenerate, allowing the $2n$ two-component 
eigenstates to be paired to form $n$ four-component Dirac
eigenstates.\\
$\bullet$ If $M_D$ = 0, the left- and right-handed components
decouple, yielding $n$ left-handed Majorana eigenstates with
standard model interactions. 
  
There are interesting physical effects associated with these 
limits.  Dirac neutrinos can have magnetic dipole, electric
dipole (CP and T violating), and anapole (P violating) moments,
as well as nonzero charge radii.  Majorana neutrinos can have
anapole moments but only transition magnetic and electric
dipole moments.  I mentioned in the previous lecture that 
transition magnetic moments were quite interesting in the 
context of MSW effects, as well as the use of $M_R$ in the
seesaw mechanism.
  
\subsection{Red Giants and Helium Burning}
We now consider the evolution off the main sequence of a solar-like
star, with a mass above half a solar mass.
As the hydrogen burning in the core progresses to the point that
no more hydrogen is available, the stellar core consists of the
ashes from this burning, $^4$He.  The star then goes through an 
interesting evolution:\\
$\bullet$ With no further means of producing energy, the core
slowly contracts, thereby increasing in temperature as gravity
does work on the core.\\
$\bullet$ Matter outside the core is still hydrogen rich, and
can generate energy through hydrogen burning.  Thus burning
in this shell of material supports the outside layers of the
star.  Note as the core contracts, this matter outside the
core also is pulled deeper into the gravitational potential.
Furthermore, the shell H burning continually adds more mass to the core.
This means the burning in the shell must intensify to generate
the additional gas pressure to fight gravity.  The shell also
thickens as this happens, since more hydrogen is above the
burning temperature.\\
$\bullet$ The resulting increasing gas pressure causes the outer
envelope of the star to expand by a larger factor, up to a 
factor of 50.  The increase in radius more than compensates for
the increased internal energy generation, so that a cooler
surface results.  The star reddens.  Thus this class of
star is named a red supergiant.\\
$\bullet$ This evolution is relatively rapid, perhaps a few
hundred million years: the dense core requires large energy
production.  The helium core is supported
by its degeneracy pressure, and is characterized by densities
$\sim 10^6$ g/cm$^3$.  This stage ends when the
core reaches densities and temperatures that allow helium burning
through the reaction
\begin{equation}
\alpha + \alpha + \alpha \rightarrow ^{12}C + \gamma .
\end{equation}
As this reaction is very temperature dependent (see below),
the conditions for ignition are very sharply defined.
This has the consequence that the core mass at the helium flash point
is well determined. \\
$\bullet$  The onset of helium burning produces a new source of
support for the core.  The energy release elevates the temperature
and the core expands: He burning, not electron degeneracy, now 
supports the core.  The burning shell and envelope have moved
outward, higher in the gravitational potential.  Thus shell
hydrogen burning slows (the shell cools) because less gas pressure
is needed to satisfy hydrostatic equilibrium.  All of this
means the evolution of the star has now slowed: the red giant
moves along the ``horizontal branch", as interior temperatures
slowly elevate much as in the main sequence.

The 3$\alpha$ process depends on some rather interesting nuclear
physics.  The first interesting ``accident" involves the near
degeneracy of the $^8$Be ground state and two separated
$\alpha$s: The $^8$Be $0^+$ ground state is just 92 keV above
the 2$\alpha$ threshold.  The measured width of the $^8$Be
ground state is 2.5 eV, which corresponds to a lifetime of
\begin{equation}
\tau_m \sim 2.6 \cdot 10^{-16} \mathrm{s} . 
\end{equation}
One can compare this number to the typical time for one $\alpha$
to pass another.  The red giant core temperature is $T_7 \sim 10
\rightarrow E \sim 8.6$ keV.  Thus v/c $\sim$ 0.002.  So the
transit time is
\begin{equation}
\tau \sim {d \over v} \sim {5f \over 0.002} {1 \over
3 \cdot 10^{10} \mathrm{cm/sec}} {10^{-13} \mathrm{cm} \over
f} \sim 8 \cdot 10^{-21} \mathrm{s} . 
\end{equation}
This is more than five orders of magnitude shorter than 
$\tau_m$ above.  Thus when a $^8$Be nucleus is produced, it lives
for a substantial time compared to this naive estimate.

To quantify this, we calculate the flux-averaged cross section
assuming resonant capture
\begin{equation}
\langle \sigma v \rangle = ( {2 \pi \over \mu k T} )^{3/2}
{\Gamma \Gamma \over \Gamma} e^{-E_r/kT} 
\end{equation}
where $\Gamma$ is the 2$\alpha$ width of the $^8$Be ground 
state.  This is the cross section for the
$\alpha+\alpha$ reaction to form the compound nucleus then 
decay by $\alpha + \alpha$.  But since there is only one 
channel, this is clearly also the result for producing the
compound nucleus $^8$Be.

By multiplying the rate/volume for producing $^8$Be by the
lifetime of $^8$Be, one gets the number of $^8$Be nuclei
per unit volume
\begin{eqnarray}
N(Be) &=& {N_\alpha N_\alpha \over 1 + \delta_{\alpha \alpha}}
\langle \sigma v \rangle \tau_m \nonumber \\
&=& {N_\alpha N_\alpha \over 1 + \delta_{\alpha \alpha}}
\langle \sigma v \rangle {1 \over \Gamma} \nonumber \\
 &=&{N_\alpha^2 \over 2} ({2 \pi \over \mu k T})^{3/2} 
e^{-E_r/kT} . 
\end{eqnarray}
Notice that the concentration is {\it independent} of $\Gamma$.
So a small $\Gamma$ is not the reason we obtain a substantial
buildup of $^8$Be.  This is easily seen: if the width is small,
then the production rate of $^8$Be goes down, but the lifetime
of the nucleus once it is produced is longer.  The two effects
cancel to give the same $^8$Be concentration.  One sees that
the significant $^8$Be concentration results from two effects:
1) $\alpha+\alpha$ is the only open channel and 2) the resonance
energy is low enough that some small fraction of the
$\alpha+\alpha$ reactions have the requisite energy.
As $E_r = 92$ keV, $E_r/kT$ = 10.67/$T_8$
(where $T_8$ is the temperature in $10^8$K) so that
\begin{equation}
N(Be) = N_\alpha^2 T_8^{-3/2} e^{-10.67/T_8} (0.94 \cdot
10^{-33} \mathrm{cm^3} ). 
\end{equation}
So plugging in typical values of $N_\alpha \sim 1.5 \cdot 
10^{28}$/cm$^3$ (corresponding to $\rho_\alpha \sim 10^5$
g/cm$^3$) and $T_8 \sim$ 1 yields
\begin{equation}
 {N(^8Be) \over N(\alpha)} \sim 3.2 \times 10^{-10} . 
\end{equation}

Salpeter suggested that this concentration would then allow
$\alpha + ^8$Be$ \rightarrow ^{12}$C to take place.  Hoyle then
argued that this reaction would not be fast enough to produce
significant burning unless it was also resonant.  Now the 
mass of $^8$Be + $\alpha$, relative to $^{12}$C,is 7.366 MeV, and each nucleus has
$J^\pi = 0^+$.  Thus s-wave capture would require a $0^+$
resonance in $^{12}$C at $\sim$ 7.4 MeV.  No such state was 
then known, but a search by Cook, Fowler, Lauritsen, and
Lauritsen revealed a $0^+$ level at 7.644 MeV, with decay
channels $^8$Be$ + \alpha$ and $\gamma$ decay to the $2^+$
4.433 level in $^{12}$C.  The parameters are
\begin{equation}
 \Gamma_\alpha \sim 8.9 \mathrm{eV} ~~~~~~~~
\Gamma_\gamma \sim 3.6 \cdot 10^{-3} \mathrm{eV} . 
\end{equation}
The resonant cross section formula gives
\begin{equation}
 r_{48} = N_8 N_\alpha ({2 \pi \over \mu kT})^{3/2}
{\Gamma_\alpha \Gamma_\gamma \over \Gamma} e^{-E_r/kT}. 
\end{equation}
Plugging in our previous expression for $N(^8$Be) yields
\begin{equation}
r_{48} = N_\alpha^3 T_8^{-3} e^{-42.9/T_8}
(6.3 \cdot 10^{-54} \mathrm{cm^6/s}).
\end{equation}
If we denote by $\omega_{3 \alpha}$ the decay rate of an $\alpha$ 
in our plasma, then
\begin{eqnarray}
 \omega_{3 \alpha} &=& 3 N_\alpha^2 T_8^{-3} 
e^{-42.9/T_8} (6.3 \cdot 10^{-54} \mathrm{cm^6/sec}) \nonumber \\
 &=& ({N_\alpha \over 1.5 \cdot 10^{28}/\mathrm{cm}^3})^2
(4.3 \cdot 10^3/\mathrm{sec}) T_8^{-3} e^{-42.9/T_8} . 
\end{eqnarray}

Now the energy release per reaction is 7.27 MeV.
Thus we can calculate the energy produced per gram, $\epsilon$:
\begin{eqnarray}
 \epsilon &=& \omega_{3 \alpha} {7.27 \mathrm{MeV} \over 3}
{1.5 \cdot 10^{23} \over \mathrm{g}} \nonumber \\ 
 &=& (2.5 \cdot 10^{21} \mathrm{erg/g~sec}) ({N_\alpha \over
1.5 \cdot 10^{28}/\mathrm{cm}^3})^2 T_8^{-3} 
e^{-42.9/T_8} . 
\end{eqnarray}
We can evaluate this at a temperature of $T_8 \sim$ 1 to find
\begin{equation}
 \epsilon \sim (584 \mathrm{ergs/g~sec}) ({N_\alpha \over
1.5 \cdot 10^{28}/\mathrm{cm}^3})^2 . 
\end{equation}
Typical values found in stellar calculations are in good agreement
with this: typical red giant energy production is $\sim$ 100
ergs per gram per second.

To get a feel for the temperature sensitivity of this process,
we can do a Taylor series expansion, finding
\begin{equation}
 \epsilon(T) \sim ({T \over T_o})^{40} N^2_\alpha . 
\end{equation}
This steep temperature dependence is the reason the He flash
is delicately dependent on conditions in the core.\\

\subsection{Neutrino Magnetic Moments and He Ignition}
Prior to the helium flash, the degenerate He core radiates
energy largely by neutrino pair emission.  The process is
the decay of a plasmon --- which one can think of as a photon
``dressed" by electron-hole excitations --- thereby acquiring 
an effective mass of about 10 keV.  The photon couples to
a neutrino pair through a electron particle-hole pair that
then decays into a $Z_o \rightarrow \nu \bar{\nu}$. 

If this cooling is somehow enhanced, the degenerate helium core 
would be kept cooler, and would not ignite at the normal
time.  Instead it would continue to grow until it overcame
the enhanced cooling to reach, once again, the ignition 
temperature. 

One possible mechanism for enhanced cooling is a neutrino
magnetic moment.  Then the plasmon could directly couple to
a neutrino pair.  The strength of this coupling would 
depend on the size of the magnetic moment.

A delay in the time of He ignition has several observable
consequences, including changing the ratio of red giant to
horizontal branch stars.  Thus, using the standard theory of
red giant evolution, investigators have attempted to determine
what size of magnetic moment would produce unacceptable 
changes in the astronomy.  The result is a limit~\cite{raff} on the
neutrino magnetic moment of
\begin{equation}
 \mu_{ij} \lsim 3 \cdot 10^{-12} \mathrm{electron~Bohr~magnetons} 
\end{equation}
as was mentioned earlier.
This limit is more than two orders of magnitude more stringent
than that from direct laboratory tests. 

This example is just one of a number of such constraints that
can be extracted from similar stellar cooling arguments.
The arguments above, for example, can be repeated for 
neutrino electric dipole moments.  More interesting, it
can be repeated for axion emission from red giants.  
Axions, the pseudoGolstone bosons associate with the solution
of the strong CP problem suggested by Peccei and Quinn,
are very light and can be produced radiatively within the
red giant by the Compton process, by the Primakoff process
off nuclei, or by emission from low-lying nuclear levels,
such at from the 14 keV transition in $^{57}$Fe.  
The net result is that axions of mass above a few eV are
excluded; if axions have a direct coupling to electrons, so
that the Compton process off electrons operates, the 
constraint is considerably tighter. 

A similar argument can be formulated for supernova cooling.
During SN1987A the neutrino burst detected by IMB and by
Kamiokande was consistent with cooling on a timescale of 
about 4 seconds.  Thus any process cooling
the star more efficiently than neutrino emission would
have shortened this time, while also reducing the flux in 
neutrinos.  Large Dirac neutrino masses allow trapped 
neutrinos to scatter into sterile right-handed states.
Right-handed neutrinos, lacking standard model interactions,
would then escape the star (provided they do not scatter
back into interacting left-handed states).
Unfortunately the upper bounds imposed on the neutrino mass
are quite model dependent, ranging over (1-25) keV. 

The supernova cooling argument can also be repeated for
axions.  The window of sensitive runs from 1 eV (above this
mass they are more strongly coupled than neutrinos, and thus
cannot compete with neutrino cooling)
to about 0.01 eV (below this mass they are too weakly coupled
to be produced on the timescale of supernova cooling).
It is interesting that the supernova and red giant cooling
limit on axions nearly meet: a small window may still exist
around a few eV if the axion has no coupling to electrons. 

\section{Supernovae, Supernova Neutrinos, and Nucleosynthesis}
Consider a massive star, in excess of 10 solar masses, burning
the hydrogen in its core under the conditions of hydrostatic
equilibrium.  When the hydrogen is exhausted, the core contracts
until the density and temperature are reached where 3$\alpha \rightarrow
^{12}$C can take place.  The He is then burned to exhaustion.
This pattern (fuel exhaustion, contraction, and ignition of the 
ashes of the previous burning cycle) repeats several times,
leading finally to the explosive burning of $^{28}$Si to Fe.
For a heavy star, the evolution is rapid: the star has to work
harder to maintain itself against its own gravity, and therefore
consumes its fuel faster.  A 25 solar mass star would go through
all of these cycles in about 7 My, with the final explosive Si 
burning stage taking a few days.  The result is an 
``onion skin" structure of the precollapse star 
in which the star's history can be read by looking at the 
surface inward: there are concentric shells of H, $^4$He,
$^{12}$C, $^{16}$O and $^{20}$Ne, $^{28}$Si, and $^{56}$Fe
at the center. 

\subsection{The Explosion Mechanism~\protect\cite{mezz}}
The source of energy for this evolution is nuclear binding energy.
A plot of the nuclear binding energy $\delta$ as a function of nuclear
mass shows that the minimum is achieved at Fe.  In a scale
where the $^{12}$C mass is picked as zero:
\begin{center}
$^{12}$C~~~~~$\delta$/nucleon = 0.000 MeV \\
$^{16}$O~~~~~$\delta$/nucleon = -0.296 MeV \\
$^{28}$Si~~~~$\delta$/nucleon = -0.768 MeV \\
$^{40}$Ca~~~~$\delta$/nucleon = -0.871 MeV \\
$^{56}$Fe~~~~$\delta$/nucleon = -1.082 MeV \\
$^{72}$Ge~~~~$\delta$/nucleon = -1.008 MeV \\
$^{98}$Mo~~~~$\delta$/nucleon = -0.899 Mev
\end{center}
Thus once the Si burns to produce Fe, there is no further source
of nuclear energy adequate to support the star.  So as the last
remnants of nuclear burning take place, the core is largely
supported by degeneracy pressure, with the energy generation rate
in the core being less than the stellar luminosity.  The core
density is about 2 $\times 10^9$ g/cc and the temperature is
kT $\sim$ 0.5 MeV. 

Thus the collapse that begins with the end of Si burning is
not halted by a new burning stage, but continues.  As gravity
does work on the matter, the collapse leads to a rapid heating
and compression of the matter.  As the nucleons in Fe are bound 
by about 8 MeV, sufficient heating can release $\alpha$s and a few
nucleons.  At the same time, the electron chemical potential is
increasing.  This makes electron capture on nuclei and any free
protons favorable,
\begin{equation}
 e^- + p \rightarrow \nu_e + n. 
\end{equation}
Note that the chemical equilibrium condition is
\begin{equation}
 \mu_e + \mu_p = \mu_n + \langle E_\nu \rangle. 
\end{equation}
Thus the fact that neutrinos are not trapped plus the rise in
the electron Fermi surface as the density increases, lead to
increased neutronization of the matter.  The escaping neutrinos carry
off energy and lepton number.  Both the electron capture and
the nuclear excitation and disassociation take energy out of the electron gas,
which is the star's only source of support.  This means that 
the collapse is very rapid.  Numerical simulations find that 
the iron core of the star ($\sim$ 1.2-1.5 solar mases) collapses
at about 0.6 of the free fall velocity.

In the early stages of the infall the $\nu_e$s readily escape.
But neutrinos are trapped when a 
density of $\sim$ 10$^{12}$g/cm$^3$ is reached. 
At this point the neutrinos begin to scatter off the matter through
both charged current and coherent neutral current processes.  The
neutral current neutrino scattering off nuclei is particularly
important, as the scattering cross section is off the total nuclear
weak charge, which is approximately the 
neutron number.  This process transfers very little energy because
the mass energy of the nucleus is so much greater than the
typical energy of the neutrinos.  But momentum is exchanged.  
Thus the neutrino ``random walks" out of the star.  When the
neutrino mean free path becomes sufficiently short, the ``trapping
time" of the neutrino begins to exceed the time scale for the
collapse to be completed.  This occurs at a density of about
10$^{12}$ g/cm$^3$, or somewhat less than 1\% of nuclear density.
After this point, the energy released by further gravitational
collapse and the star's remaining lepton number are trapped
within the star. 

If we take a neutron star of 1.4 solar masses and a radius of
10 km, an estimate of its binding energy is
\begin{equation}
 {G M^2 \over 2R} \sim 2.5 \times 10^{53} \mathrm{ergs}. 
\end{equation}
Thus this is roughly the trapped energy that will later be radiated in neutrinos. 

The trapped lepton fraction $Y_L$ is a crucial parameter in the
explosion physics: a higher trapped $Y_L$ leads to a larger
homologous core, a stronger shock wave, and easier passage of
the shock wave through the outer core, as will be discussed 
below.  Most of the 
lepton number loss of an infalling mass element occurs as it
passes through a narrow range of densities just before trapping.
The reasons for this are relatively simple: on dimensional 
grounds weak rates in a plasma 
go as $T^5$, where T is the temperature.  Thus the electron capture rapidly turns on as
matter falls toward the trapping radius, and lepton number loss is
maximal just prior to trapping.  Inelastic neutrino reactions
have an important effect on these losses, as the
coherent trapping cross section goes as $E_\nu^2$ and is thus 
least effective for the lowest energy neutrinos.  As these
neutrinos escape, inelastic reactions repopulate the low
energy states, allowing the neutrino emission to continue.

The velocity of sound in matter rises with increasing density.
The inner homologous core, with a mass $M_{HC} \sim 0.6-0.9
$ solar masses, is that part of the iron core where the sound
velocity exceeds the infall velocity.  This allows any pressure
variations that may develop in the homologous core during infall
to even out before the collapse is completed.  As a result, the
homologous core collapses as a unit, retaining its density
profile.  That is, if nothing were to happen to prevent it, 
the homologous core would collapse to a point. 

The collapse of the homologous core continues until nuclear
densities are reached.  As nuclear matter is rather incompressible ($\sim$ 200 MeV/f$^3$),
the nuclear equation of state is effective in halting the collapse:
maximum densities of 3-4 times nuclear are reached, e.g.,
perhaps $6 \cdot 10^{14}$ g/cm$^3$.  The innermost shell of matter
reaches this supernuclear density first, rebounds, sending a 
pressure wave out through the homologous core.  This wave
travels faster than the infalling matter, as the homologous 
core is characterized by a sound speed in excess of the infall
speed.  Subsequent shells follow.  The resulting series of pressure
waves collect near the sonic point (the edge of the homologous
core).  As this point reaches nuclear density and comes to
rest, a shock wave breaks out and begins its traversal of the 
outer core. 

Initially the shock wave may carry an order of magnitude more energy
than is needed to eject the mantle of the star (less than 10$^{51}$
ergs).  But as the shock wave travels through the outer iron core,
it heats and melts the iron that crosses the shock front, at a 
loss of $\sim$ 8 MeV/nucleon.  The enhanced electron capture 
that occurs off the free protons left in the wake of the shock,
coupled with the sudden reduction of the neutrino opacity of
the matter (recall $\sigma_{coherent} \sim N^2$), greatly 
accelerates neutrino emission.  This is another energy loss.
[Many numerical models predict a strong ``breakout" burst of 
$\nu_e$s in the few milliseconds required for the shock wave to
travel from the edge of the homologous core to the neutrinosphere
at $\rho \sim 10^{12}$ g/cm$^3$ and $r \sim 50$ km.
The neutrinosphere is the term from the neutrino 
trapping radius, or surface of last scattering.]  The summed losses
from shock wave heating and neutrino emission are comparable to 
the initial energy carried by the shock wave.  Thus most 
numerical models fail to produce a successful ``prompt"
hydrodynamic explosion. 

Most of the attention in the past decade focused on two explosion
scenarios.  In the prompt mechanism described above, the shock wave
is sufficiently strong to survive the passage of the outer iron
core with enough energy to blow off the mantle of the star.
The most favorable results were achieved with smaller stars
(less than 15 solar masses) where there is less overlying iron,
and with soft equations of state, which produce a more compact
neutron star and thus lead to more energy release.  In part
because of the lepton number loss problems discussed earlier,
now it is widely believed that this mechanism fails for all but
unrealistically soft nuclear equations of state. 

The delayed mechanism begins with a failed hydrodynamic explosion;
after about 0.01 seconds the shock wave stalls at a radius of
200-300 km.  It exists in a sort of equilibrium, gaining energy
from matter falling across the shock front, but loosing energy
to the heating of that material.  However, after perhaps 0.5
seconds, the shock wave is revived due to neutrino heating of 
the nucleon ``soup" left in the wake of the shock.  This heating
comes primarily from charged current reactions off the nucleons
in that nucleon gas; quasielastic scattering also contributes.
This high entropy radiation-dominated gas may reach two MeV in temperature.
The pressure exerted by this gas helps to 
push the shock outward. It is important to note
that there are limits to how effective this neutrino energy 
transfer can be: if matter is too far from the core, the coupling
to neutrinos is too weak to deposite significant energy.  If too
close, the matter may be at a temperature (or soon reach a temperature)
where neutrino emission cools the matter as fast or faster than
neutrino absorption heats it.  The term
``gain radius" is used to describe the region where
useful heating is done. 

This subject is still controversial and unclear.  The
problem is numerically challenging, forcing modelers
to handle the difficult hydrodynamics of a shock wave; the
complications of the nuclear equation of state at densities not
yet accessible to experiment; modeling in two or three dimensions;
handling the slow diffusion of neutrinos; etc.  Not all of these
aspects can be handled reasonably at the same time, even with
existing supercomputers.  Thus there is considerable disagreement
about whether we have any supernova model that succeeds in
ejecting the mantle. 

However the explosion proceeds, there is agreement that 99\% 
of the 3 $\cdot 10^{53}$ ergs released in the collapse is 
radiated in neutrinos of all flavors.  The time scale over 
which the trapped neutrinos leak out of the protoneutron star
is about three seconds.
Through most of their migration out of the protoneutron
star, the neutrinos are in flavor equilibrium
\begin{equation}
 \mathrm{e.g.},~~ \nu_e + \bar{\nu}_e \leftrightarrow \nu_\mu + \bar{\nu}_\mu. 
\end{equation}
As a result, there is an approximate equipartition of energy
among the neutrino flavors.  After weak decoupling, the $\nu_e$s 
and $\bar{\nu_e}$s remain in equilibrium with the matter for
a longer period than their heavy-flavor counterparts, due to 
the larger cross sections for scattering off electrons and 
because of the charge-current reactions
\begin{eqnarray}
 \nu_e + n &&\leftrightarrow p + e^- \nonumber \\ 
 \bar{\nu_e} + p &&\leftrightarrow n + e^+. 
\end{eqnarray}
Thus the heavy flavor neutrinos decouple from deeper within the star, 
where temperatures are higher.  Typical 
calculations yield 
\begin{equation}
 T_{\nu_\mu} \sim T_{\nu_\tau} \sim 8 \mathrm{MeV} ~~~~
 T_{\nu_e} \sim 3.5 \mathrm{MeV}~~~~T_{\bar{\nu_e}} \sim 4.5 \mathrm{MeV}. 
\end{equation}
The difference between the $\nu_e$ and $\bar{\nu_e}$ temperatures
is a result of the neutron richness of the matter, which enhances
the rate for charge-current reactions of the $\nu_e$s, thereby keeping them coupled
to the matter somewhat longer. 

This temperature hierarchy is crucially important to nucleosynthesis
and also to
possible neutrino oscillation scenarios.  The three-flavor MSW
level-crossing diagram is shown in Fig. 10.  One very popular
scenario attributes the solar neutrino problem to $\nu_\mu \leftrightarrow \nu_e$
transmutation; this means that a second crossing with a $\nu_\tau$
could occur at higher density.  It turns out plausible seasaw
mass patterns suggest a $\nu_\tau$ mass on the order of a few eV,
which would be interesting cosmologically.  The second crossing
would then occur outside the neutrino sphere, that is, after
the neutrinos have decoupled and have fixed spectra with the
temperatures given above.  Thus a $\nu_e \leftrightarrow \nu_\tau$ oscillation
would produce a distinctive $T \sim 8$ MeV spectrum of $\nu_e$s.
This has dramatic consequences for terrestrial detection and 
for nucleosynthesis in the supernova. 

\begin{figure}[htb]
\psfig{bbllx=1.0cm,bblly=4.0cm,bburx=18cm,bbury=18.5cm,figure=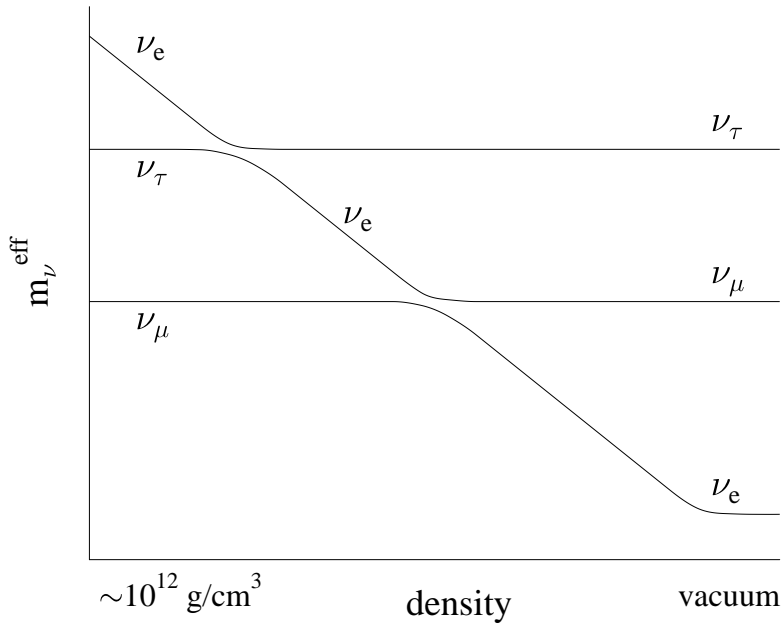,height=3.5in}
\caption{Three-flavor neutrino level-crossing diagram.  One 
popular scenario associates the solar neutrino problem with
$\nu_e \leftrightarrow \nu_\mu$ oscillations and predicts 
a cosmologically interested massive $\nu_\tau$ with 
$\nu_e \leftrightarrow \nu_\tau$ oscillations near the
supernova neutrinosphere.}
\end{figure}
  
\subsection{The Neutrino Process~\protect\cite{nupro}}
Core-collapse supernovae are one of the
major engines driving galactic chemical evolution, producing 
and ejecting the metals that enrich our galaxy.  The discussion
of the previous section described the hydrostatic evolution of
a presupernova star in which large quantities of the most
abundant metals (C, O, Ne, ...) are synthesized and later
ejected during the explosion.  During the passage of the
shock wave through the star's mantle, temperature of $\sim (1-3) \cdot 10^9$K and
are reached in the silicon, oxygen, and neon shells.  This
shock wave heating induces $(\gamma,\alpha) \leftrightarrow 
(\alpha,\gamma)$ and related reactions that generate a 
mass flow toward highly bound nuclei, resulting in the
synthesis of iron peak elements as well as less abundant
odd-A species.  Rapid neutron-induced reactions are thought
to take place in the high-entropy atmosphere just above
the mass cut, producing about half of the heavy elements 
above A $\sim$ 80.  This is the subject of the Sec. 4.3.
Finally, the $\nu$-process described below is responsible 
for the synthesis of rare species such as $^{11}$B and $^{19}$F.
This process involves the response of nuclei at momentum transfers
where the allowed approximation is no longer valid.  Thus we
will use the $\nu$-process in this section to illustrate some of
the relevant nuclear physics. 

One of the problems -- still controversial -- that may be connected
with the neutrino process is
the origin of the light elements Be, B and Li, elements which are
not produced in sufficient amounts in the big bang or in any of
the stellar mechanisms we have discussed.
The traditional explanation has been cosmic ray spallation interactions
with C, O, and N in the interstellar medium.  In this picture,
cosmic ray protons collide with C at relatively high energy,
knocking the nucleus apart.  So in the debris one can find 
nuclei like $^{10}$B, $^{11}$B, and $^7$Li.

But there are some problems with this picture.  First of all,
this is an example of a secondary mechanism: the interstellar
medium must be enriched in the C, O, and N to provide the 
targets for these reactions.  Thus cosmic ray spallation must 
become more effective as the galaxy ages.  The 
abundance of boron, for example, would tend to grow 
quadratically with metalicity, since the rate of production
goes linearly with metalicity.  But
observations, especially recent measurements with the 
HST, find a linear growth~\cite{timmes} in the boron abundance. 

A second problem is that the spectrum of cosmic ray protons
peaks near 1 GeV, leading to roughly comparable production of the
two isotopes $^{10}$B and $^{11}$B.  That is, while it takes 
more energy to knock two nucleons out of carbon than one, this
difference is not significant compared to typical cosmic ray
energies.  More careful studies
lead to the expectation that the abundance ratio
of $^{11}$B to $^{10}$B might be $\sim$ 2.  In nature, it is
greater than 4.

Fans of cosmic ray spallation have offered solutions to these
problems, e.g., similar reactions occurring in the atmospheres
of nebulae involving lower energy cosmic rays.  
As this suggestion was originally stimulated by the observation of nuclear
$\gamma$ rays from Orion, now retracted, some of the motivation
for this scenario has evaporated.  Here I 
focus on an alternative explanation, synthesis via neutrino spallation.

Previously we described the allowed
Gamow-Teller (spin-flip) and Fermi weak interaction operators.  These are
the appropriate operators when one probes the nucleus at
a wavelength -- that is, at a size scale -- where the nucleus
responds like an elementary particle.  We can then 
characterize its response by its macroscopic quantum numbers,
the spin and charge.  On the other hand, the nucleus is a
composite object and, therefore, if it is probed at shorter
length scales, all kinds of interesting radial excitations will
result, analogous to the vibrations of a drumhead.  
For a reaction like neutrino scattering off a nucleus, the
full operator involves the additional factor
\begin{equation}
e^{i \vec{k} \cdot \vec{r}} \sim 1 + i \vec{k} \cdot \vec{r} 
\end{equation}
where the expression on the right is valid if the magnitude of
$\vec{k}$ is not too large.  Thus the full charge operator 
includes a ``first forbidden" term
\begin{equation}
 \sum_{i=1}^A \vec{r}_i \tau_3(i) 
\end{equation}
and similarly for the spin operator
\begin{equation}
 \sum_{i=1}^A [\vec{r}_i \otimes \vec{\sigma}(i)]_{J=0,1,2} \tau_3(i). 
\end{equation}
These operators generate collective radial excitations,
leading to the so-called ``giant resonance" excitations in nuclei.
The giant resonances are typically at an excitation energy of
20-25 MeV in light nuclei.  One important property is that these
operators satisfy a sum rule (Thomas-Reiche-Kuhn) of the form
\begin{equation}
 \sum_f | \langle f | \sum_{i=1}^A r(i) \tau_3(i) | i \rangle |^2
\sim {N Z \over A} \sim {A \over 4} 
\end{equation}
where the sum extends over a complete set of final nuclear states.
These first-forbidden operators tend to dominate the cross sections
for scattering the high energy supernova neutrinos ($\nu_{\mu}$s 
and $\nu_\tau$s), with $E_\nu \sim$ 25 MeV, off light nuclei.
From the sum rule above, it follows that nuclear cross sections per
target {\it nucleon} are roughly constant. 

The E1 giant dipole mode described above is depicted qualitatively
in Fig. 11a.  This description, which corresponds to an early model
of the giant resonance response by Goldhaber and Teller, 
involves the harmonic oscillation of the proton and neutron 
fluids against one another.  The restoring force for small 
displacements would be linear in the displacement and 
dependent on the nuclear symmetry energy.  There is a natural
extension of this model to weak interactions, where axial
excitations occur.  For example, one can envision a mode
similar to that of Fig. 11a where
the spin-up neutrons and spin-down protons oscillate against
spin-down neutrons and spin-up protons, the spin-isospin mode
of Fig. 11b.  This mode is one that arises in a
simple SU(4) extension of the Goldhaber-Teller model,
derived by assuming that the nuclear force is spin and isospin
independent, at the same excitation energy as the E1 mode.
In full, the Goldhaber-Teller model predicts a degenerate 15-dimensional supermultiplet of 
giant resonances, each obeying sum rules analogous to 
the TRK sum rule.  While more sophisticated descriptions of the
giant resonance region are available, of course, this crude
picture is qualitatively accurate. 
  
\begin{figure}[htb]
\psfig{bbllx=0.3cm,bblly=2.8cm,bburx=13cm,bbury=11.5cm,figure=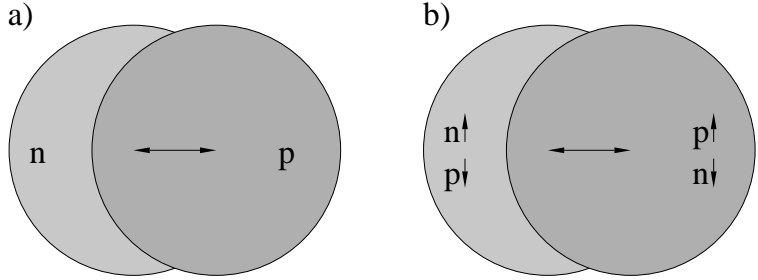,height=1.9in}
\caption{Schematic illustration of a) the E1 giant dipole mode
familiar from electromagnetic interactions and b) a spin-isospin
giant dipole mode associated with the first-forbidden weak
axial response.}
\end{figure}
  
This nuclear physics is important to the $\nu$-process.
The simplest example of $\nu$-process nucleosynthesis involves the Ne shell
in a supernova.  Because of the first-forbidden contributions,
the cross section for inelastic neutrino scattering to the 
giant resonances in Ne is $\sim 3 \cdot 10^{-41}$ cm$^2$/flavor
for the more energetic heavy-flavor neutrinos.
This reaction
\begin{equation}
 \nu + A \rightarrow \nu' + A^* 
\end{equation}
transfers an energy typical of giant resonances, $\sim$ 20 MeV.
A supernova releases about 3 $\times 10^{53}$ ergs
in neutrinos, which converts to about $4 \times 10^{57}$ heavy
flavor neutrinos.  The Ne shell in a 20 M$_\odot$ star has
at a radius $\sim$ 20,000 km.  Thus the neutrino fluence through
the Ne shell is
\begin{equation}
 \phi \sim { 4 \cdot 10^{57} \over 4 \pi (20,000 \mathrm{km})^2 }
\sim 10^{38}/\mathrm{cm}^2. 
\end{equation}
Thus folding the fluence and cross section,
one concludes that approximately 1/300th of the Ne nuclei interact.

This is quite interesting since the astrophysical origin of $^{19}$F
had not been understood.  The only stable isotope of fluorine,
$^{19}$F has an abundance
\begin{equation}
 {^{19}\mathrm{F} \over ^{20}\mathrm{Ne}} \sim {1 \over 3100}. 
\end{equation}
This leads to the conclusion that the fluorine 
found in toothpaste was
created by neutral current neutrino reactions deep inside some
ancient supernova. 

The calculation of the final $^{19}$F/$^{20}$Ne ratio is 
more complicated than the simple 1/300 ratio given above: \\
$\bullet$ When Ne is excited by $\sim$ 20 MeV through inelastic
neutrino scattering, it breaks up in two ways
\begin{eqnarray}
 ^{20}\mathrm{Ne}(\nu,\nu')^{20}\mathrm{Ne}^* &&\rightarrow ^{19}\mathrm{Ne} + n 
\rightarrow ^{19}\mathrm{F} + e^+ + \nu_e + n \nonumber \\
 ^{20}\mathrm{Ne}(\nu,\nu')^{20}\mathrm{Ne}^* &&\rightarrow ^{19}\mathrm{F}
+ p 
\end{eqnarray}
with the first reaction occurring half as frequently as the 
second.  As both channels lead to $^{19}$F, we have correctly
estimated the instantaneous abundance ratio in the Ne shell of
\begin{equation}
 {^{19}\mathrm{F} \over ^{20}\mathrm{Ne}} \sim {1 \over 300}. 
\end{equation}
$\bullet$ We must also address the issue of whether the produced $^{19}$F
survives.  In the first 10$^{-8}$ sec the coproduced neutrons in
the first reaction react via
\begin{equation}
^{15}\mathrm{O}(n,p)^{15}\mathrm{N}~~^{19}\mathrm{Ne}(n,\alpha)^{16}\mathrm{O}~~
^{20}\mathrm{Ne}(n,\gamma)^{21}\mathrm{Ne}~~^{19}\mathrm{Ne}(n,p)^{19}\mathrm{F} 
\end{equation}
with the result that about 70\% of the $^{19}$F produced via
spallation of neutrons is then immediate destroyed, primarily
by the $(n,\alpha)$ reaction above.  In the next $10^{-6}$ sec
the coproduced protons are also processed
\begin{equation}
 ^{15}\mathrm{N}(p,\alpha)^{12}\mathrm{C}~~^{19}\mathrm{F}(p,\alpha)^{16}\mathrm{O}~~
^{23}\mathrm{Na}(p,\alpha)^{20}\mathrm{Ne} 
\end{equation}
with the latter two reactions competing as the primary proton
poisons.  This makes an important prediction: stars with high Na
abundances should make more F, as the $^{23}$Na acts as a proton
poison to preserve the produced F.\\
$\bullet$ Finally, there is one other destruction mechanism, the
heating associated with the passage of the shock wave.  It
turns out the the F produced prior to shock wave passage can
survive if it is in the outside half of the Ne shell.  The reaction
\begin{equation}
 ^{19}\mathrm{F}(\gamma,\alpha)^{15}\mathrm{N} 
\end{equation}
destroys F for peak explosion temperatures exceeding $1.7 \cdot 10^9$K.
Such a temperature is produced at the inner edge of the Ne 
shell by the shock wave heating, but not at the outer edge.

If all of this physics in handled is a careful network code that
includes the shock wave heating and F production both before and
after shock wave passage, the following are the results:
 \[ \begin{array}{cc} \underline{[^{19}\mathrm{F}/^{20}\mathrm{Ne}]/
[^{19}\mathrm{F}/^{20}\mathrm{Ne}]_\odot} & \underline{T_{\mathrm{heavy}~\nu} \mathrm{(MeV)}} \\
0.14 & 4 \\ 0.6 & 6 \\ 1.2 & 8 \\ 1.1 & 10 \\ 1.1 & 12 \end{array} \]
where the abundance ratio in the first column has been normalized
to the solar value.
One sees that the attribution of F to the neutrino process argues
that the heavy flavor $\nu$ temperature must be greater than 6 MeV,
a result theory favors.  One also sees that F cannot be overproduced
by this mechanism: although the instantaneous production of F
continues to grow rapidly with the neutrino temperature, too
much F results in its destruction through the $(p,\alpha)$
reaction, given a solar abundance of the competing proton poison
$^{23}$Na.  Indeed, this illustrates an odd quirk: although 
in most cases the neutrino process is a primary mechanism, one needs
$^{23}$Na present to produce significant F. Thus in this case the neutrino
process is a secondary mechanism. 

While there are other significant neutrino process products ($^7$Li,
$^{138}$La, $^{180}$Ta, $^{15}$N ...), the most important 
product is $^{11}$B, produced by spallation off carbon.
A calculation by Timmes et al. [18] found that the combination of
the neutrino process, cosmic ray spallation and big-bang 
nucleosythesis together can explain the evolution of the light
elements.  The neutrino process, which produces a great deal 
of $^{11}$B but relatively little $^{10}$B, combines with the
cosmic ray spallation mechanism to yield the observed
isotope ratio.  Again, one prediction of this picture is that
early stars should be $^{11}$B rich, as the neutrino process 
is primary and operates early in our galaxy's history; the
cosmic ray production of $^{10}$B is more recent.
There is hope that HST studies will soon be able to descriminate
between $^{10}$B and $^{11}$B: as yet this has not been done. 

\subsection{The r-process}
Beyond the iron peak nuclear Coulomb barriers become so high
that charged particle reactions become ineffective, leaving
neutron capture as the mechanism responsible for producing
the heaviest nuclei.
If the neutron abundance is modest,
this capture occurs in such a way that each newly synthesized
nucleus has the opportunity to $\beta$ decay, if it is energetically
favorable to do so.  Thus weak equilibrium is maintained within
the nucleus, so that synthesis is along the path of stable 
nuclei.  This is called the s- or slow-process.  However a
plot of the s-process in the (N,Z) plane reveals that this
path misses many stable, neutron-rich nuclei that are known to
exist in nature.  This suggests that another mechanism is at
work, too.  Furthermore, the abundance peaks found in nature 
near masses A $\sim$ 130 and A $\sim$ 190, which mark the closed
neutron shells where neutron capture rates and $\beta$ decay
rates are slower, each split into two subpeaks.  One set of subpeaks
corresponds to the closed-neutron-shell numbers N $\sim$ 82
and N $\sim$ 126, and is clearly associated with the s-process.
The other set is shifted to smaller N, $\sim$ 76 and $\sim$ 116,
respectively, and is suggestive of a much more explosive
neutron capture environment where neutron capture can be
rapid. 
  
This second process is the r- or rapid-process, characterized by: \\
$\bullet$ The neutron capture is fast compared to $\beta$ decay rates. \\
$\bullet$ The equilibrium maintained within a nucleus is established by $(n,\gamma) \leftrightarrow
(\gamma,n)$: neutron capture fills up the available bound levels in
the nucleus until this equilibrium sets in.  The new Fermi level
depends on the temperature and the relative $n/\gamma$ abundance.\\
$\bullet$ The nucleosynthesis rate is thus controlled by the $\beta$
decay rate: each $\beta^-$ capture coverting n $\rightarrow$ p 
opens up a hole in the neutron Fermi sea, allowing another neutron
to be captured. \\
$\bullet$ The nucleosynthesis path is along exotic, neutron-rich
nuclei that would be highly unstable under normal laboratory conditions. \\
$\bullet$ As the nucleosynthesis rate is controlled by the $\beta$
decay, mass will build up at nuclei where the $\beta$ decay rates
are slow.  It follows, if the neutron flux is reasonable steady 
over time so that equilibrated mass flow is reached, that the
resulting abundances should be inversely proportional to these
$\beta$ decay rates. 
  
Let's first explore the $(n,\gamma) \leftrightarrow (\gamma,n)$
equilibrium condition, which requires that the rate for $(n,\gamma)$
balances that for $(\gamma,n)$ for an average nucleus.
So consider the formation cross section
\begin{equation}
 A + n \rightarrow (A+1) + \gamma . 
\end{equation}
This is an exothermic reaction, as the neutron drops into the
nuclear well.  Our averaged cross section, assuming a resonant
reaction (the level density is high in heavy nuclei) is
\begin{equation}
\langle \sigma v \rangle_{(n,\gamma)} = 
\left( {2 \pi \over \mu kT} \right)^{3/2} {\Gamma_n \Gamma_\gamma
\over \Gamma} e^{-E/KT} 
\end{equation}
where E $\sim$ 0 is the resonance energy,
and the $\Gamma$s are the indicated partial and total widths.
Thus the rate per unit volume is
\begin{equation}
r_{(n,\gamma)} \sim N_n N_A \left( {2 \pi \over \mu kT} \right)^{3/2}
{\Gamma_n \Gamma_\gamma \over \Gamma}
\end{equation}
where $N_n$ and $N_A$ are the neutron and nuclear number densities
and $\mu$ the reduced mass.
This has to be compared to the $(\gamma,n)$ rate. 

The $(\gamma,n)$ reaction requires the photon number density in
the gas.  This is given by the Bose-Einstein distribution
\begin{equation}
N(\epsilon) = {8 \pi \over c^3 h^3} {\epsilon^2 d \epsilon
\over e^{\epsilon/kT} -1} .
\end{equation}
The high-energy tail of the normalized distribution can thus
be written
\begin{equation}
 \sim {1 \over N_\gamma \pi^2} \epsilon^2 e^{-\epsilon/kT} d \epsilon 
\end{equation}
where in the last expression we have set $\hbar = c = 1$. 

Now we need the resonant cross section in the $(\gamma,n)$ 
direction.  For photons the wave number is proportional to
the energy, so
\begin{equation}
\sigma_{(\gamma,n)} = {\pi \over \epsilon^2}
{\Gamma_\gamma \Gamma_n \over (\epsilon-E_r)^2 + (\Gamma/2)^2 } .
\end{equation}
As the velocity is c =1,
\begin{equation}
\langle \sigma v \rangle = {1 \over \pi^2 N_\gamma}
\int_0^\infty \epsilon^2 e^{-\epsilon/kT} d \epsilon
{\pi \over \epsilon^2} {\Gamma_\gamma \Gamma_n \over 
(\epsilon-E_r)^2 +(\Gamma/2)^2} . 
\end{equation}
We evaluate this in the usual way for a sharp resonance,
remembering that the energy integral over just the denominator
above (the sharply varying part) is $2 \pi/ \Gamma$
\begin{equation}
 \sim {\Gamma_\gamma \Gamma_n \over N_\gamma} e^{-E_r/kT}
{2 \over \Gamma} . 
\end{equation}
So that the rate becomes
\begin{equation}
r_{(\gamma,n)} \sim 2 N_{A+1} {\Gamma_\gamma \Gamma_n
\over \Gamma} e^{-E_r/kT} .  
\end{equation}
Equating the $(n,\gamma)$ and $(\gamma,n)$ rates and taking
$N_A \sim N_{A-1}$ then yields
\begin{equation}
N_n \sim {2 \over (\hbar c)^3} \left( {\mu c^2 kT \over
2 \pi} \right)^{3/2} e^{-E_r/kT} 
\end{equation}
where the $\hbar$s and $c$s have been properly inserted to give
the right dimensions.  Now $E_r$ is esssentially the binding
energy.  So plugging in the conditions $N_n \sim 3 \times 10^{23}$/cm$^3$
and $T_9 \sim 1$, we find that the binding energy is 
$\sim$ 2.4 MeV.  Thus neutrons are bound by about 30 times $kT$,
a value that is still small compared to a typical
binding of 8 MeV for a normal nucleus.  (In this calculation
I calculated the neutron reduced mass assuming a nuclear target
with A=150.) 

The above calculation fails to count spin states for the photons
and nuclei and is thus not quite correct.  But it makes the
essential point: the r-process involves very exotic species
largely unstudied in any terrestrial laboratory.  It is good
to bear this in mind, as in the following section we will 
discuss the responses of such nuclei to neutrinos.  Such responses
thus depend on the ability of theory to extrapolate responses
from known nuclei to those quite unfamiliar. 

The path of the r-process is along neutron-rich nuclei, 
where the neutron Fermi sea is just $\sim$ (2-3) MeV away from
the neutron drip line (where no more bound neutron levels exist).
After the r-process finishes (the neutron exposure ends)
the nuclei decay back to the valley of stability by $\beta$
decay.  This can involve some neutron spallation ($\beta$-delayed
neutrons) that shift the mass number A to a lower value.
But it certainly involves conversion of neutrons into protons,
and that shifts the r-process peaks at N $\sim$ 82 and 126
to a lower N, off course.  This effect is clearly seen in the
abundance distribution: the r-process peaks are shifted to
lower N relative to the s-process peaks.  This is the origin of the 
second set of ``subpeaks" mentioned at the start of the section. 

It is believed that the r-process can proceed to very heavy
nuclei (A $\sim$ 270) where it is finally ended by $\beta$-delayed
and n-induced fission, which feeds matter back into the
process at an A $\sim$ A$_{max}$/2.  Thus there may be important
cycling effects in the upper half of the r-process distribution.
  
What is the site(s) of the r-process?  This has been debated 
many years and still remains a controversial subject:\\
$\bullet$ The r-process requires exceptionally explosive conditions 
\begin{center}
$\rho$(n) $\sim 10^{20}$ cm$^{-3}$~~~T $\sim 10^9$K~~~t $\sim$ 1s.
\end{center}
$\bullet$ Both primary and secondary sites proposed. 
Primary sites are those not requiring preexisting metals.
Secondary sites are those where the neutron capture occurs
on preexisting s-process seeds.\\
$\bullet$ Suggested primary sites include the
the neutronized atmosphere above the proto-neutron star in
a Type II supernova, neutron-rich jets produced in supernova
explosions or in neutron star mergers, inhomogeneous big
bangs, etc. \\
$\bullet$ Secondary sites, where $\rho$(n) can be lower for 
successful synthesis, include the He and C zones in Type II
supernovae, the red giant He flash, etc.

The balance of evidence favors a primary site, so one requiring
no preenrichment of heavy s-process metals.  Among the evidence: \\
  
\noindent
1) HST studies of very-metal-poor halo stars: 
The most important evidence are the recent HST measurements of 
Cowan, Sneden et al.~\cite{sneden} of very metal-poor stars ([Fe/H] $\sim$ -1.7 to -3.12)
where an r-process distribution very much like that of our sun
has been seen for Z $\gsim$ 56.  Furthermore, in these stars
the iron content is variable.  This suggests that the ``time
resolution" inherent in these old stars is short compared to
galactic mixing times (otherwise Fe would be more constant).
The conclusion is that the r-process material in these stars
is most likely from one or a few local supernovae.  The fact
that the distributions match the solar r-process (at least 
above charge 56) strongly suggests that there is some kind of
unique site for the r-process: the solar r-process distribution
did not come from averaging over many different kinds of
r-process events.  Clearly the fact that these old stars are
enriched in r-process metals also strongly argues for a 
primary process: the r-process works quite well in an
environment where there are few initial s-process metals.\\

\noindent
2) There are also fairly good theoretical arguments that a primary
r-process occurring in a core-collapse supernova might be
viable~\cite{hotbub}.  First, galactic chemical evolution studies indicate that 
the growth of r-process elements in the galaxy is consistent 
with low-mass Type II supernovae in rate and distribution.
More convincing is the fact that modelers have shown that the
conditions needed for an r-process (very high neutron densities,
temperatures of 1-3 billion degrees) might be realized in a
supernova.  The site is the last material expelled from the 
supernova, the matter just above the mass cut.  When
this material is blown off the star initially, it is a very
hot neutron-rich, radiation-dominated gas containing neutrons
and protons, but an excess of the neutrons.  As it expands
off the star and cools, the material first goes through
a freezeout to $\alpha$ particles, a step that essentially
locks up all the protons in this way.
Then the $\alpha$s interact through reactions like 
\begin{eqnarray}
 \alpha + \alpha +\alpha &&\rightarrow ^{12}C  \nonumber \\
 \alpha + \alpha + n &&\rightarrow ^9Be \nonumber
\end{eqnarray}
to start forming heavier nuclei.  Note, unlike the big bang,
that the density is high enough to allow such three-body 
interactions to bridge the mass gaps at A = 5,8.  The
$\alpha$ capture continues up to heavy nuclei,
to A $\sim$ 80, in the network calculations.  
The result is a small number of ``seed" nuclei,
a large number of $\alpha$s, and excess neutrons.  These 
neutrons preferentially capture on the heavy seeds to
produce an r-process.  Of course, what is necessary is to
have $\sim$ 100 excess neutrons per seed in order to 
successfully synthesize heavy mass nuclei.  Some of the
modelers find conditions where this almost happens. 
  
There are some very nice aspects of this site: the amount of
matter ejected is about 10$^{-5} - 10^{-6}$ solar masses,
which is just about what is needed over the lifetime of the
galaxy to give the integrated r-process metals we see,
taking a reasonable supernova rate.  But there are also
a few problems, especially the fact that with calculated entropies
in the nucleon soup above the proto-neutron star, neutron fractions
appear to be too low to produce a successful A $\sim$ 190 peak.
There is some interesting recent work invoking neutrino oscillations
to cure this problem: charge current reactions on free protons
and neutrons determine the n/p ratio in the gas.  Then, for example, an oscillation
of the type $\nu_e \rightarrow \nu_{\mathrm{sterile}}$ can alter this
ratio, as it would turn off the $\nu_e$s that destroy neutrons
by charged-current reactions.  Unfortunately, 
a full discussion of such possibilities would take
us too far afield today. 

The nuclear physics of the r-process tells us that the synthesis
occurs when the nucleon soup is in the temperature range of
(3-1) $\cdot 10^9$K, which, in the hot bubble r-process described above, corresponds to a freezeout radius of
(600-100) km and a time $\sim$ 10 seconds after core collapse.
The neutrino fluence after freezeout (when the temperature
has dropped below 10$^9$K and the r-process stops) is then $\sim$
(0.045-0.015) $\cdot 10^{51}$ ergs/(100km). 
Thus, after completion of the r-process, the newly synthesized
material experiences an intense flux of neutrinos.
This brings up the question of whether the neutrino flux could
have any effect on the r-process.  

\subsection{Neutrinos and the r-process~\protect\cite{qian}}
Rather than describe the exotic effects of neutrino oscillations
on the r-process, mentioned briefly above, we will examine
standard-model effects that are nevertheless quite interesting.
The nuclear physics of this section -- neutrino-induced neutron
spallation reactions -- is also relevant to recently proposed
supernova neutrino observatories such as OMNIS and LAND.
In contrast to our first discussion of the $\nu$-process in
Sec. 4.2, it is apparent that neutrino effects could be much
larger in the hot bubble r-process: the synthesis
occurs {\it much} closer to the star than our Ne radius of
20,000 km: estimates are 600-1000 km.  The r-process is completed
in about 10 seconds (when the temperature drops to about 
one billion degrees), but the neutrino flux is still significant
as the r-process freezes out.  The net result is that the
``post-processing" neutrino fluence - the fluence that can
alter the nuclear distribution after the r-process is completed -
is about 100 times larger than that responsible for fluorine
production in the Ne zone.  Recalling that 1/300 of the nuclei
in the Ne zone interacted with neutrinos, and remembering that
the relevant neutrino-nucleus cross sections scale as A, one
quickly sees that the probability of a r-process nucleus 
interacting with the neutrino flux is approximately unity.

Because the hydrodynamic conditions of the r-process are highly
uncertain, one way to attack this problem is to work backward
in time.  We know the final r-process distribution (what nature
gives us) and we can calculate neutrino-nucleus interactions
relatively well.  Thus from the observed r-process distribution
(including neutrino postprocessing) we can work backward to
find out what the r-process distribution looked like at the
point of freezeout.  In Figs. 12 and 13, the ``real" r-process
distribution - that produced at freezeout - is given by the 
dashed lines, while the solid lines show the effects of the
neutrino postprocessing for a particular choice of fluence. 
The nuclear physics input into these calculations is precisely
that previously described: GT and first-forbidden cross sections,
with the responses centered at excitation energies consistent
with those found in ordinary, stable nuclei, taking into
account the observed dependence on $|N-Z|$. 

\begin{figure}[htb]
\psfig{bbllx=-2.0cm,bblly=4.5cm,bburx=18cm,bbury=23.0cm,figure=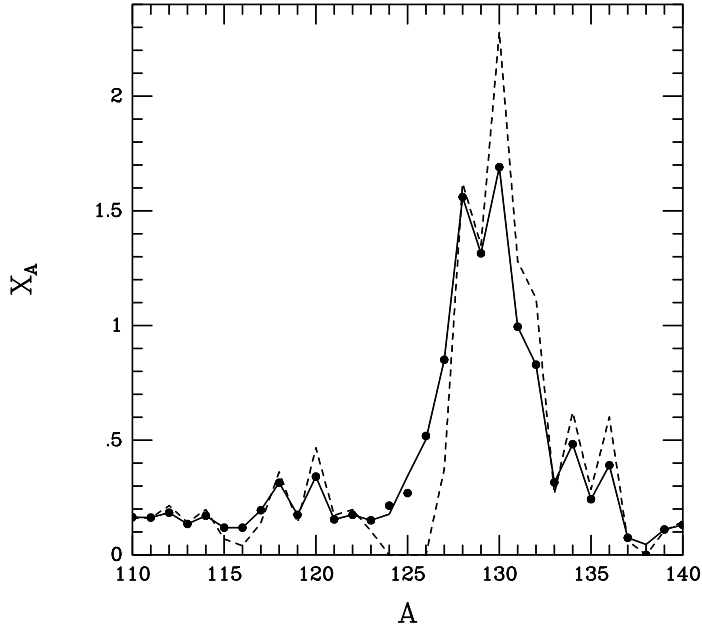,height=3.5in}
\caption{Comparison of the r-process distribution that would 
result from the freezeout abundances near the A $\sim$ 130 
mass peak (dashed line) to that where the effects of neutrino
postprocessing have been include (solid line).  The fluence 
has been fixed by assuming that the A = 124-126 abundances
are entirely due to the $\nu$-process.}
\end{figure}
  
One important aspect of the figures is that the mass shift is
significant.  This has to do with the fact that a 20 MeV 
excitation of a neutron-rich nucleus allows multiple neutrons
( $\sim$ 5) to be emitted.  
(Remember we found that the binding energy of the last neutron
in an r-process neutron-rich nuclei was about 2-3 MeV under
typical r-process conditions.)  The second thing to notice is that
the relative contribution of the neutrino process is particularly
important in the ``valleys" beneath the mass peaks: the reason
is that the parents on the mass peak are abundant, and the
valley daughters rare.  In fact, it follows from this that the neutrino
process effects can be dominant for precisely seven
isotopes (Te, Re, etc.) lying in these valleys.  Furthermore
if an appropriate neutrino fluence is picked, these isotope
abundances are produced perfectly (given the abundance errors).
The fluences are
\begin{eqnarray}
     \mathrm{N} &=& 82~ \mathrm{peak}~~~~~0.031 \cdot 10^{51} \mathrm{ergs/(100km)^2/flavor} \nonumber \\
     \mathrm{N} &=& 126~ \mathrm{peak}~~~~0.015 \cdot 10^{51} \mathrm{ergs/(100km)^2/flavor}, \nonumber
\end{eqnarray}
values in fine agreement with those that would be found
in a hot bubble r-process.  So this is circumstantial but 
significant evidence that the material near the mass cut of 
a Type II supernova is the site of the r-process: there is a
neutrino fingerprint. 

\begin{figure}[htb]
\psfig{bbllx=-2.0cm,bblly=4.5cm,bburx=18cm,bbury=23.0cm,figure=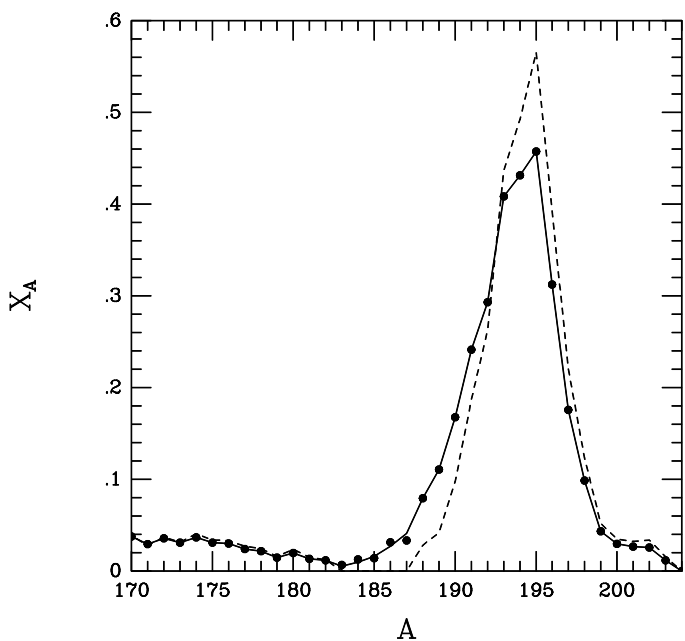,height=3.5in}
\caption{As in Fig. 12, but for the A $\sim$ 195 mass peak.
The A = 183-187 abundances are entirely attributed to the 
$\nu$-process.}
\end{figure}

\subsection{Neutrino Oscillations and the r-process}
For the usual seesaw pattern of neutrino masses and a
cosmological interesting $\nu_\tau$ (i.e., a heavy neutrino 
with a mass in the neighborhood of 10 eV), the full MSW 
pattern is shown in Fig. 10.  If the $\nu_e - \nu_\mu$
crossing is responsible for the solar neutrino problem,
a second crossing, $\nu_e - \nu_\tau$, is expected at a density
large compared to that of the solar core, but small compared
to the location of the supernova neutrinosphere 
($\sim 10^{12}$ g/cm$^3$).  For a very large range of mixing
angles, this crossing is adiabatic and thus leads to 
$\nu_e \leftrightarrow \nu_\tau$ conversion.  These spectra
thus change identities, leading to an anomalously hot $\nu_e$
flux from a Type II supernova. 
  
As the $\nu$-nucleon cross section is proportional to E$_\nu^2$,
the reaction 
$\nu_e$ + n $\rightarrow$ e$^-$ + p is enhanced, while 
$\bar{\nu}_e$ + p $\rightarrow$ e$^+$ + n is unchanged.
For a rather extensive range of $\nu_e \leftrightarrow \nu_\tau$
mixing angles and $\delta m^2$, this crossing then destroys the
r-process: the hotter $\nu_e$s drive the matter proton rich \cite{fuller}.
Thus, if one accepts this location as the site of the r-process,
very strong constraints on cosmologically interesting $\nu_\tau$s
are obtained.  These limits are truly remarkable for their
sensitivity to small mixing angles, extended to 
$\sin^2 2 \theta \sim 10^{-5}$ for neutrino mass differences
above a few eV$^2$.

I thank A. S. Brun and R. E. Shrock for helpful comments,
and Paul Langacker for his able organization of the 1998 TASI
summer school.
This work was supported in part by the US Department of Energy.

\end{document}